\documentclass[12pt,reqno, a4paper]{amsart}
\usepackage{graphicx}
\usepackage{fancyhdr}
\usepackage{enumerate}
\usepackage{amsmath}

\usepackage{setspace}
\usepackage{amsthm}
\usepackage{mathabx}
\usepackage{amssymb}
\usepackage{relsize}
\usepackage{pdfsync}
\usepackage[colorlinks=true,linkcolor=blue,citecolor=magenta]{hyperref}

\usepackage[normalem]{ulem}

\usepackage{xcolor}

\usepackage{calligra}

\usepackage{tikz}

\usepackage{changebar}
\usepackage{pdfsync}
\usetikzlibrary{decorations.pathmorphing,patterns}
\usetikzlibrary{calc,arrows}

\usepackage{pgf}

\usepackage{mathtools}

\DeclarePairedDelimiter\floor{\lfloor}{\rfloor}

\usetikzlibrary{automata}
\usetikzlibrary{positioning}

\tikzset{
    state/.style={
           rectangle,
           rounded corners,
           draw=black, very thick,
           minimum height=2em,
           inner sep=2pt,
           text centered,
           },
}

\usepackage{MnSymbol}

\usepackage{xcolor}

\usepackage{changebar}
\usepackage{pdfsync}

\usepackage[normalem]{ulem}

\makeindex
\newtheorem{theorem}{Theorem}[section]
\newtheorem{lemma}[theorem]{Lemma}
\newtheorem{proposition}[theorem]{Proposition}
\newtheorem{corollary}[theorem]{Corollary}

\newtheorem{remark}[theorem]{Remark}

\newcommand{\cal}{\mathcal}

\newcommand\bbR{{\mathbb R}}
\newcommand\bbZ{{\mathbb Z}}

\newcommand\al{\alpha}
\newcommand\la{\lambda}

\newcommand\Om{{\Omega}}
\newcommand\R{{\mathbb R}}

\newcommand \ga{\gamma}
\newcommand \om{\omega}
 
\newcommand \si{\sigma}
\newcommand{\bbT}{\mathbb T}

\renewcommand{\ge}{\geqslant}
\renewcommand{\le}{\leqslant}
\newcommand{\dd  }{\mathrm{d}}

\renewcommand{\hat}{\widehat}
\renewcommand{\tilde}{\widetilde}
\renewcommand{\bar}{\overline}
\numberwithin{equation}{section}

\begin{document}

\title{Convergent Power Series for Anharmonic  Chain with Periodic Forcing}

\author{Pedro L. Garrido}
 \address{Pedro L. Garrido\\Universidad de Granada\\Granada Spain} 
\email{{\tt garrido@onsager.ugr.es}}

 \author{Tomasz Komorowski}
 \address{Tomasz Komorowski, Institute of Mathematics,
   Polish Academy Of Sciences, Warsaw, Poland.} 
\email{{\tt tkomorowski@impan.pl}}

\author{Joel L. Lebowitz}
\address{Joel L. Lebowitz, Departments of Mathematics and Physics,  Rutgers University}
\email{\tt lebowitz@math.rutgers.edu}

 \author{Stefano Olla}
 \address{Stefano Olla, CEREMADE,
   Universit\'e Paris-Dauphine, PSL Research University \\
 \and Institut Universitaire de France\\ \and
GSSI, L'Aquila}
  \email{\tt olla@ceremade.dauphine.fr}

%\date{\today {\bf File: {\jobname}.tex.}} 

\begin{abstract}

We study the propagation of energy in one-dimensional anharmonic chains subject to a periodic,
localized forcing. For the purely harmonic case, forcing frequencies outside the
linear spectrum produce exponentially localized responses, preventing
equi-distribution of energy per degree of freedom.
We extend this result to anharmonic perturbations with bounded second derivatives and
boundary dissipation, proving that for small perturbations and non-resonant forcing,
the dynamics converges to a periodic stationary state with energy exponentially
localized uniformly in the system size. The perturbed periodic state is described by a convergent
power type expansion in the {strength} of the anharmonicity.
This excludes chaoticity induced by  anharmonicity,  
independently of the size of the system. Our perturbative scheme can
also be applied in higher dimensions.
%Our perturbative scheme also applies to disordered perturbations,
% connecting with Anderson localization. Open questions remain for resonant forcing
% and unbounded nonlinearities such as in the Fermi-Pasta-Ulam-Tsingou problem.

\end{abstract}

\thanks{P.G.  acknowledges the support of 
  the Project I+D+i Ref.No. PID2023-149365NB-I00, funded by
  MICIU/AEI/10.13039/501100011033/ and by ERDF/EU, T.K  acknowledges the support of the
  NCN grant 2024/53/B/ST1/00286} 
 \keywords{anharmonic chain, periodic force, spectrum of the infinite
   harmonic chain, perturbation series, stability.}
  \subjclass[2000]{80A19,80M22,82C22,82C70,70J35}

\maketitle

%\tableofcontents

\section{Introduction}
\label{sec:intro}

The approach to equilibrium which involves energy sharing among
degrees of freedom of large systems  with non-linear dynamics is one of the
central problems in statistical mechanics. This motivated the numerical experiment
in the seminal work of Fermi-Pasta-Ulam-Tsingou \cite{fpu}, which was perhaps the
first application of digital computers to a non-military scientific goal.
They considered a one-dimensional anharmonic chain of $N = 32$ oscillators, initially
prepared in a periodic low-energy configuration. Surprisingly, the simulated evolution
failed to show thermalization, i.e. an equidistribution of the energy among all other
frequencies generated by the nonlinearity of the dynamics. The results of these
simulations remain a source of investigation and a challenge to theoreticians
\cite{Lepri_7,Gallavotti_2,cit1_3,Lepri_12}. Later studies repeated the simulations
for much larger systems and at higher energies, showing that thermalization can indeed
occur. However, no complete analytical results in this direction are known.

% According to the KAM theorem, nontrivial integrals of motion — that is, conserved
% quantities beyond energy and momentum — may persist in finite systems that are small
% perturbations of linear dynamics. It is nevertheless believed that, for \emph{generic}
% nonlinearities, these conserved quantities \emph{disappear} in the large-size limit
% ($N \to \infty$), in the sense that the corresponding invariant surfaces fold densely
% into the microcanonical ones. There are, however, important counterexamples of
% completely integrable extended systems, such as the Toda lattice, which show that
% the situation is more subtle, and that thermalization may strongly depend on the
% specific nature of the nonlinearities.

In this article we approach the problem from a different perspective. Instead of
starting the dynamics of the anharmonic chain in a given periodic configuration, we
apply a periodic force of period $\theta$ at a single site of the chain.
For the linear dynamics it is known that only frequencies within the spectral interval
can propagate through the system. In fact, in \cite{Garrido} we proved that if the
forcing frequency lies outside the frequency spectrum of the system, the perturbation
induced by the forcing remains exponentially localized. This is not surprising, since
linear dynamics is completely integrable and cannot thermalize.

Here we consider anharmonic perturbations of the linear dynamics, with potentials
having bounded second derivatives. Dissipation is applied at the boundaries, ensuring
the existence of a periodic stationary state for any system size $N$.
We prove that, for sufficiently small perturbations and if the forcing frequency is
outside the resonance spectrum of the unperturbed linear system, the dynamics,
starting from any initial configuration, converges in time to a periodic stationary
state with energy exponentially localized uniformly in $N$. In particular, this implies
that no thermalization occurs under these circumstances. This result is obtained via a
rigorous perturbative scheme around the purely harmonic case. Although the method
can be extended to   higher dimension and   infinite systems, we refrain from doing so here for simplicity.

An open question remains as to what happens when the forcing frequency lies within the
resonance interval of the unperturbed linear system, or when the anharmonicity is
unbounded (as in the FPUT case); see some numerical results in the companion article
\cite{Garrido2}.

% Our perturbative expansion is robust enough to also cover disordered (or random)
% perturbations. In particular, perturbing the deterministic ordered harmonic chain with
% a random harmonic still yields the same result, in agreement with Anderson localization.

As far as we know, there are no other analytical results of this kind for non-linear
dynamics that are uniform in the system size. There has also been considerable work
on transport properties of anharmonic chains, both for systems coupled to thermal
reservoirs and for systems driven by external periodic forces but at positive temperature,
see \cite{Lepri_3,Lepri_11,Dolg,Prem_3,Gallavotti_3,Lepri_4,Lepri_9,Lepri_10,Lepri_5,Lepri_6,Lepri_1,Lepri_2,Lepri_8}
and references therein. 

Computer simulations on similar systems, where a quartic
anharmonicity is added to a pinned harmonic potential, reveal a rich variety of steady
states emerging as the driving parameters of the external force are varied
\cite{Prem_1,Prem_2,Prem_3,Prem_4,Prem_5,Prem_6,Geniet_2,Geniet_1}.

\section{Microscopic dynamics}
\label{sec:micro}

The
configuration of the system is given by
\begin{equation}
  \label{eq:1}
  (\mathbf q, \mathbf p) =
  (q_{-N}, \dots, q_N, p_{-N}, \dots, p_N) \in \Omega_N:= \R^{2N+1}
  \times \R^{2N+1}. 
\end{equation}
{We denote  $\bbZ_N:=\{-N,\ldots,N\}$.}
The microscopic dynamics of the chain are given by the
forced Hamiltonian system with the Hamiltonian given by
{\begin{equation}
\label{010503-25}
{\cal H}_N({\bf q},{\bf p})=\sum_{x\in\bbZ_N}\left[\frac{p_x^2}2 +
\frac12 (q_{x}-q_{x-1})^2 +\frac{\om_0^2 q_x^2}{2}+\nu\Big( V(q_x)+
U(q_x-q_{x-1})\Big) \right],\end{equation}}
friction on both endpoints of the lattice interval
and a driving periodic force at $x=0$. This yields
\begin{equation} 
\label{012805-24x} %HZ100
\begin{aligned}
   \ddot    q_x(t;\nu) &=  \Delta q_x(t;\nu)-\om_0^2 q_x(t;\nu)  -\ga
  \big(\delta_{-N}(x)+ \delta_{N}(x)\big) \dot q_x(t;\nu)\\
  &
 -\nu\Big( V'(q_x(t;\nu)) -\nabla U'\big(q_x(t;\nu)-q_{x-1}(t;\nu)\big) \Big)+ {\cal F}(t/\theta)\delta_{x,0} ,\quad
  x\in \bbZ_N,
\end{aligned} \end{equation}
with {strictly positive  parameters $\gamma, \,\om_0,\, \theta$.}
  The Neumann laplacian $\Delta $ and the discrete gradient $\nabla$ are defined as
\begin{equation}
  \label{nmn}
  \Delta f_x=f_{x+1}+f_{x-1}-2f_x,\quad \nabla
  f_x=f_{x+1}-f_x,\quad  x\in\bbZ_N,
\end{equation}
with the boundary condition
\begin{equation}
  \label{eq:2}
  f_{N+1}=f_N,\quad f_{-N-1}=f_{-N}.
\end{equation}

{The assumption that $\gamma>0$ is our standing hypothesis and, unless otherwise stated, is in
force throughout the paper.}

%the parameters $\gamma,\om_0,\theta>0$ and, by convention, $q_{-N-1}=q_{-N}$ and $q_{N+1}=q_N$. 

We let the force be of the form
\begin{equation}
  \label{forcing-g}
  \begin{split}
   & {\cal F}(t/\theta)=\sum_{m\in\bbZ}\hat{\rm F}_me^{im\om t},\quad
   \mbox{where } 
   \hat{\rm F}_{0}=0\quad\mbox{and}\\
   &
   \hat{\rm F}_m^\star =\hat{\rm F}_{-m}\quad\mbox{and}\quad
   0<\sum_{m\in\bbZ}\big(m|\hat{\rm F}_m|\big)^2<+\infty.
    \end{split}
  \end{equation}
  Here
  \begin{equation}
    \label{om-theta}
    \om=\frac{2\pi}{\theta}.
    \end{equation}

      In the special instance  $N=0$, $\bbZ_0:=\{0\}$ the above model
     can be formulated as follows
\begin{equation} 
\label{012301-25}
\begin{aligned}
    \ddot    q_0(t;\nu) &=  -\om_0^2 q_0(t;\nu)  -2\ga
   \dot q_0(t;\nu)
   -\nu V'(q_0(t;\nu)) + {\cal F}(t/\theta)  
 \end{aligned} \end{equation}
{and it  constitutes a case of its own interest, see ref. \cite{Dyakonov_1,Dyakonov_2} }. {We will discuss it in Section \ref{sec:case-n=0}.}
      
%Combining both equations of \eqref{eq:HZ100} (resp. \eqref{012301-25})
%we conclude that the atom   displacement satisfies a non-linear  discrete wave equation
%\begin{equation} 
%\label{012805-24x}
%\begin{split} &  \ddot    q_x(t;\nu) =  (\Delta -\om_0^2) q_x(t;\nu)
%   -\ga
 %  (\delta_{x,-N}+
%  \delta_{x,N})  \dot q_x(t;\nu) \\
%   &
%   -\nu\Big(
%   V'(q_x(t;\nu)) -\nabla U'\big(q_x(t;\nu)-q_{x-1}(t;\nu)\big)\Big)
% + {\cal F}(t/\theta)\delta_{x,0} ,\quad
%   x\in \bbZ_N
%   \end{split}
%\end{equation}
%for $N=1,2,\ldots$ and,
%in the case $N=0$ 
% \begin{equation} 
%\label{012805-24x0}
%\ddot    q_0(t;\nu) =
%  -\om_0^2 q_0(t;\nu)-2\ga \dot q_0(t;\nu)
%   -\nu
%   V'(q_0(t;\nu)) 
% + {\cal F}(t/\theta).
%\end{equation}

We consider the case where the
non-quadratic part of the pinning and interacting potentials
$V(q)$, and $ U(r)$ respectively, are  of $C^2$ class of regularity
\footnote{{Our argument actually requires only $C^1$ smoothness and
  the bounded Lipschitz assumption on the derivative.}}
with bounded second derivatives, i.e.  
\begin{equation}
  \label{Vp1z}
 \| V''\|_{\infty}+\| U''\|_{\infty}<+\infty. %,\quad V'(0)=U'(0).
\end{equation}
Here for a given  function $G:\bbR\to\bbR$ we denote by
\begin{equation}
  \label{lip}
\| G\|_{\infty}:=\sup_{q }|G(q)|.
\end{equation}
Examples of such  potentials are furnished
by 
\begin{equation}
 V(q)= \frac{q^{2n}}{1+\alpha q^{2n}}\quad\mbox{for some }
 \,\al>0,\label{pot1}
\end{equation}
or
\begin{equation}
 V(q)=  (\sin q)^{2n},\qquad  V(q)=  (1+\alpha q^2)^{\delta/2},\label{pot2}
\end{equation}
where {$n\ge 1$} and $\delta\in(0,2)$. Analogous examples of interaction
potentials, e.g. $U(r)=\cos r$, are also covered.

We prove that when all integer multiplicities of
$\omega=\frac{2\pi}{\theta}$ are outside the interval containing the
spectrum of the harmonic chain ${\cal
  I}:=[\omega_0,\om_u]$, with $\om_u:=\sqrt{\omega_0^2+4}$, a unique
state of the system is achieved, as $t\to+\infty$, for  {an 
arbitrary initial condition}. In addition, the state is $\theta$-periodic and given by a perturbative
  expansion in the parameter $\nu$, that  
  converges for $|\nu|<\nu_0$, uniformly in  the size of the system  $N$.
More precisely
\begin{equation}
  \label{011602-25}
\nu_0=\frac{\delta_*}{{\frak V}} .
\end{equation}
where  $\delta_*:=\inf\Big[|(m\om)^2-w^2|:\,m\in\bbZ,\,w\in
      {\cal I}\Big]$, and
  ${\frak V}=\|V''\|_\infty+3\|U''\|_\infty$.

Finding a periodic stable solution of  \eqref{012805-24x} by
a convergent perturbative series is, we believe, {unlikely for  
nonlinear anharmonicity that grows faster than quadratic,} considered in \cite{cit1_1,cit1_2,cit1_3}, and, as far as we
know, not proven for any other non-equilibrium system. 
Our results do not extend to the case where some integer multiplicity
of $\omega$ lies inside spectrum of the harmonic chain or
$\vert\nu\vert>\nu_0$.

\section{Main Results}
\label{sec-main-res}

\subsection{Definition of norms} 
To make the above statements precise we define the Euclidean norm
$$
\|f\|_{N}:=\left(\sum_{x\in\bbZ_N}f_x^2\right)^{1/2}<+\infty
$$
on the space $\bbR^{2N+1}$   of all real valued sequences
$(f_x)_{x\in\bbZ_N}$.
 Consider the space $L^2\big([0,\theta]; \bbR^{2N+1}\big)$ of all
 Borel measurable functions $F:[0,\theta]\to \bbR^{2N+1}$ equipped
with the norm
\begin{equation}
  \label{norm}
  \|\!|F\|\!|_N:=\left(\frac{1}{\theta}\int_0^\theta\|F(t)\|_{N}^2\dd t\right)^{1/2}
  \end{equation}
  for any Borel measurable $F: [0,\theta]\to\ell_2(\bbZ_N)$.  
We shall also consider  spaces
$H^k\big([0,\theta]; \bbR^{2N+1}\big)$, for $k=1,2,\ldots$, obtained
by the completion of the
space of  $C^\infty$-smooth functions $F$ in the norm
\begin{equation}
  \label{norm}
  \|\!|F\|\!|_{N,k}:=\left(\|\!|F\|\!|_N^2+\frac{1}{\theta}\sum_{\ell=1}^k\int_0^\theta\|F^{(\ell)}(t)\|_{N}^2\dd t\right)^{1/2}.
  \end{equation}
  Here $F^{(\ell)}(t)$ denotes the $\ell$-th derivative.
   {{In the following, when no ambiguity would arise,
  we drop the subscript $N$.}}

\subsection{A periodic  solution  for an anharmonic
  chain}
\label{sec3.2}

In the case when   $\gamma >0$
we state the following existence result,
whose proof is fairly routine and relies on the application of the Schauder
fixed point theorem (see Appendix  \ref{sec6}).
\begin{theorem}
  \label{thm011010-24}
 { Assume $\gamma>0$.}
  For each $\nu\in\bbR$ there exists a $\theta$-periodic solution $
  {\bf q}_{\rm p}(t;\nu)=\Big(q_{x,{\rm p}} (t;\nu)\Big)_{x\in\bbZ_N}$
  of \eqref{012805-24x}.
  % The solution is unique if $|\nu|\le \nu_0$, where $\nu_0$ is as in
  % the statement of Theorem \ref{main:thm}.
\end{theorem}
% The proof of the result is fairly routine and relies on the application of the Schauder
% fixed point theorem. We present it in Section \ref{sec6}.

\subsection{Construction of a periodic solution for an anharmonic
  chain by perturbative series}

\label{sec3.3}
% Before formulating the scheme let us introduce some notation. 
Given a sequence $f:=(f_x)_{x\in\bbZ_N}$ define 
\begin{equation}
  \label{W}
  W_x(f):=V'(f_x)-\nabla U'(f_x-f_{x-1}) .
  \end{equation}
% Suppose that $N=1,2,\ldots$. 
 %  Its time harmonics satisfy
%  \begin{equation} 
% \label{eq:HZ-0}
%   0 =  \left[\Delta  
%   +(m\om)^2-\om_0^2 -i\ga m\om(
%     \delta_{-N}(x)+
%     \delta_N(x))\right]  \tilde q_x^{(0)}(m;\nu ) 
%   +  \;      \hat{\rm  F}_m  \delta_{0}(x)
%     , \; \quad x\in\bbZ_N.
%   \end{equation}
   We consider a sequence $\Big(Q_{x,{\rm p}}^{(L)}
 (t;\nu)\Big)$, $x\in\bbZ_N$, $L=0,1,\ldots$
of functions, $\theta$-periodic in $t$, that satisfy
\begin{equation} 
\label{012805-24LL}
\begin{split} 
  \ddot    Q_{x,{\rm p}}^{(L)} (t;\nu)=  &\big(\Delta  -\om_0^2\big) Q_{x,{\rm p}}^{(L)} (t;\nu)
  -\ga(\delta_{-N}(x)+ \delta_N(x))\dot Q_{x,{\rm p}}^{(L)}(t;\nu)
   \\
   &
   -\nu  W_x\big(Q^{(L-1)}_{\rm p}(t;\nu)\big)  
  + {\cal F}(t/\theta) \delta_{0}(x),\quad
   x\in \bbZ_N.
   \end{split}
 \end{equation}
We let
\begin{equation} 
\label{012805-24R}
W_x\big(Q^{(-1)}_{\rm p}(t;\nu)\big)\equiv 0\quad\mbox{ and }\quad  Q^{(0)}_{x,{\rm p}}(t;\nu) :=q^{(0)}_{x,{\rm p}}(t;\nu),
\end{equation}
where $q_{x,{\rm p}}^{(0)}(t;\nu) = q_{x,{\rm p}}^{(0)}(t)$  is the
{unique} $\theta$-periodic solution  
of  \eqref{012805-24x} for $\nu = 0$. 
 For $L\ge 1$ we define
 \begin{equation}
   \label{eq:3}
   q^{(L)}_{x,{\rm p}}(t;\nu) := \nu^{-L} \left(Q_{x,{\rm p}}^{(L)}(t;\nu) - Q_{x,{\rm p}}^{(L-1)}(t;\nu)\right)
 \end{equation}
i.e.
 % and suppose  that   $q^{(\ell)}_{x,{\rm p}}(t;\nu)$,
% $\ell=0,1,\ldots$ are such that
\begin{equation}
  \label{010905-24}
  Q_{x,{\rm p}}^{(L)}(t;\nu)=\sum_{\ell=0}^{L}q^{(\ell)}_{x,{\rm p}}(t;\nu)\nu^\ell.
\end{equation}
A straightforward calculation (see \eqref{012805-24L} below) shows that 
$q_{x }^{(L)}(t;\nu)$, $x\in\bbZ_N$ is a $\theta$-periodic solution of 
\begin{equation} 
\label{eq:flip-lZa}
\begin{aligned}
     \ddot    q_x^{(L)}(t;\nu) &=  \Delta
   q_x^{(L)}(t;\nu)-\om_0^2 q_x^{(L)}(t;\nu)\\
   &
   -\ga\dot
   q_x^{(L)}(t;\nu)\delta_{-N}(x)-\ga\dot
    q_x^{(L)}(t;\nu)\delta_{N}(x)-   v_{x,L-1}(t;\nu)  ,\quad
  x\in \bbZ_N,
\end{aligned} \end{equation}
where
\begin{equation}
  \label{vxL}
  \begin{split}
    &v_{x,L-1}(t;\nu)=\frac{1}{\nu^{L-1}}
    \Big[W_x\big(Q^{(L-1)}_{\rm p}(t;\nu)\big)
    -W_x\big(Q^{(L-2)}_{\rm p} (t;\nu)\big)\Big].
\end{split}
\end{equation}
Note that
$
v_{x,0}(t;\nu)= W_x\big(q_{{\rm p}}^{(0)}(t;\nu) \big) .
$

{Since \eqref{eq:flip-lZa} can be treated as a harmonic chain with
 a prescribed periodic forcing,
  a standard argument, similar to the one used in Section
  \ref{sec:time-harm},
  proves that  the  scheme described above has a unique $\theta$-periodic solution.
}

% Adapting the argument made in the proof of  Proposition
% \ref{prop010309-24} we conclude that the  scheme described above
% is well defined  for a finite chain of size $2N+1$. 
% \begin{proposition}
%   \label{prop012711-24}
%   For any $\nu\in\bbR$ and $N=1,2,\ldots$  there exists
%   a unique $\theta$-periodic solution of the scheme described in
%   \eqref{012805-24LL}.  
% \end{proposition}

% The scheme described in \eqref{eq:flip-lZ}--\eqref{010905-24} can be
% also adapted to the case of an infinite lattice $\bbZ$
% (i.e. $N=+\infty$). We just drop the damping terms, i.e. those containing
% the term $\ga$, and use the lattice laplacian, instead of the Neumann
% one.

An essential assumption used in our next
result 
is that no integer multiplicity of the forcing frequency belongs to the
spectrum of the harmonic chain on the lattice $\bbZ$, with no boundary
condition, i.e.
\begin{equation}
\label{A}
m\om\not\in {\cal I}:=[\om_0, \sqrt{\om_0^2+4}]\quad\mbox{ for all }m\in\bbZ.
\end{equation}
%Here $\om_u:=\sqrt{\om_0^2+4}$.
% For $N=1,2,\ldots$ d
We recall that
  \begin{equation}
    \label{delta-m}
    \begin{split}
    &  \delta_*=\inf\Big[|(m\om)^2-w^2|:\,m\in\bbZ,\,w\in
      {\cal I}\Big],\\
      &
     { {\frak V}= \|V''\|_{\infty}+3 \|U''\|_{\infty}}.
      % \|V'\|_{\rm Lip}^2+9\|U'\|_{\rm Lip}^2+6\|U'\|_{\rm
      %   Lip}\|V'\|_{\rm Lip}.
         \end{split}
       \end{equation}
       By taking $m=0$ we observe  that $\delta_*\le \om_0^2$.
       Let
    \begin{equation}
      \label{nu0}
      \nu_0:=\frac{\delta_*}{{\frak V} }.
\end{equation}
      Our main result can be formulated as follows.
 \begin{theorem}
  \label{main:thm}
  Suppose that  $\delta_*>0$. Then,  the following
  estimate holds %(see \eqref{021602-25})
  \begin{align}
  \label{082302-24}
  \|\!|q^{(\ell+1)}_{\rm p}(\cdot;\nu)\|\!| \le
    \frac{1}{\nu_0}\|\!|q^{(\ell)}_{\rm p}(\cdot;\nu)\|\!|, \quad \nu\in\bbR,\, 
    \ell = 0,1,2,\ldots.  
  \end{align}
  In consequence,  
    for $|\nu|<\nu_0$
    the sums  $Q_{x,{\rm p}}^{(L)}(t;\nu)$, given by \eqref{010905-24}  converge
    in the $\|\cdot\|_{N,1}$-norm (see \eqref{norm}), as $L\to+\infty$.
    The convergence is uniform in $N$ and
   \begin{equation}
     \label{112302-24}
     \begin{split}
     &     \|\!|\sum_{\ell=
        L}^{+\infty}q^{(\ell)}_{\rm p} (\cdot;\nu)\nu^\ell\|\!|
  \le \frac{|\nu/\nu_0|^L \|\!| q^{(0)}_{\rm p}(\cdot;\nu)\|\!|}{1-|\nu/\nu_0|},\quad
  L=1,2,\ldots,\,|\nu| < \nu_0 .
\end{split}
\end{equation}
Consequently, for $|\nu|< \nu_0$ the series,
\begin{equation}
  \label{010401-24z}
 q_{x,{\rm p}}(t;\nu)=\sum_{\ell=0}^{+\infty}q^{(\ell)}_{x,{\rm p}}(t;\nu)\nu^\ell.
\end{equation}
 defines a $\theta$-periodic solution of \eqref{012805-24x}.
  \end{theorem}

Our  next result concerns the uniqueness of the  $\theta$-periodic
solution of \eqref{012805-24x}.
  \begin{theorem}
    \label{thm-unique}
    For $|\nu|< \nu_0$ the system
  \eqref{012805-24x} has a unique $\theta$-periodic solution.
  \end{theorem}
  Hence, for $|\nu| < \nu_0$, \eqref{010401-24z} defines the unique $\theta$-periodic solution
  of \eqref{012805-24x}.
 
   {The proofs of Theorems \ref{main:thm} and \ref{thm-unique} are
     presented in Section \ref{sec:proof-conv-pert}.}

\subsection{Global stability of the unique periodic solution in the
  anharmonic case}

 {Theorems \ref{main:thm} and \ref{thm-unique}
hold even in the absence of dissipation at the boundaries ($\gamma =
0$)    in the case of an infinite chain ($N=+\infty$).}
Of course in  {the case  $\ga=0$} the unique periodic solution has
no stability properties  {(as can be seen even for $\nu=0$)},
it is just a very special solution of \eqref{012805-24x}.
Stability holds only  when $\gamma>0$.

For $t\ge s$ denote by ${\bf q}  (t,s,{\bf q},{\bf p} ;\nu)=\big(q_x(t,s,{\bf q},{\bf p} ;\nu)\big)$
the solution of \eqref{012805-24x}
satisfying the initial condition
\begin{equation}
  \label{eq:4}
  {\bf q}  (s,s,{\bf q},{\bf p} ;\nu ) = {\bf q}, \qquad \dot{\bf q} (s,s,{\bf q},{\bf p} ;\nu ) = {\bf p}.
\end{equation}
Then $\big({\bf q}  (t,s,{\bf q},{\bf p} ;\nu ), \dot{\bf q}  (t,s,{\bf q},{\bf p} ;\nu )\big)$ tends to the
periodic solution constructed in Theorem \ref{main:thm}, as $s\to-\infty$.
\begin{theorem}
   \label{thm011111-24b}
  Suppose that $({\bf q}_{\rm p} (t ;\nu  ))$ is the unique
  periodic solution of \eqref{012805-24x}  constructed in Theorem
  \ref{main:thm} {and $\gamma >0$}. Then,  
  \begin{equation}
    \label{111010-24z}
    \lim_{s\to-\infty}   \Big( \|{\bf q}  (t,s, {\bf q},{\bf p} ;\nu )-{\bf q}_{\rm p} (t ;\nu )\|
    +\|\dot{\bf q}  (t,s, {\bf q},{\bf p};\nu )-\dot{\bf q}_{\rm p} (t ;\nu )\|\Big)=0 
  \end{equation}
for any $t\in\bbR$,   $({\bf q},{\bf p})\in\Omega_N$ and
$|\nu|<\nu_0$, with $\nu_0$ given by \eqref{nu0}.
%(resp. \eqref{nu01}) for $N=1,2,\ldots$ (resp. $N=0$).
   \end{theorem}
We present the proof of this result in Section \ref{sec5.3}.

\subsection{Work done by periodic forcing}
\label{sec:work-done-periodic}

  The work performed by the force on the
system over the period $\theta$ is given by the formula
\begin{equation}
  \label{013105-23}
 { {\cal W}_N}(\nu)=\frac{1}{\theta}\int_0^\theta {\cal F}(t/\theta)
    p_{0,{\rm p}}(t;\nu) \dd t.
  \end{equation}
  Here $p_{x,{\rm p}}(t;\nu) =\dot q_{x,{\rm p}}(t;\nu) $.
  % Recall that microscopic energy density $e_x (\mathbf q, \mathbf
 % p;\nu)$ and the Hamiltonian ${\cal H}(\mathbf q, \mathbf p;\nu)$ are
 % given by \eqref{Hal}.
We have the following identity
\begin{equation}
          \label{061705-24}
          \begin{split}
            &{\cal H}_N\big( {\bf q}_{\rm p}(t;\nu), \dot{\bf  q}_{\rm p}(t;\nu);\nu\big) - {\cal H}_N\big({\bf  q}_{\rm
        p}(0;\nu), \dot{\bf  q}_{\rm p}(0;\nu);\nu\big) \\
      &
    +\ga\Big( \int_0^{t}p_{-N, {\rm p}}^2(s;\nu)\dd
      s+\int_0^{t}p_{N, {\rm p}}^2(s;\nu)\dd s\Big)  =   \int_0^{t}p_{0, {\rm p}} (s;\nu){\cal F}\Big(\frac{s}{\theta}\Big)\dd s.
      \end{split}
    \end{equation}
    Here ${\cal H}_N(\cdot,\cdot)$ is the Hamiltonian of the system,
    see \eqref{010503-25}.
  A direct consequence of \eqref{061705-24} is the following.
\begin{theorem}
      \label{prop011705-24a}
     Suppose that {$\gamma >0$ and} $\big(q_{x,{\rm
         p}}(\cdot;\nu)\big)$ is a $\theta$-periodic solution
         of \eqref{012805-24x}.
        {Furthermore assume that the potential $U$ is such that the
          equation\footnote{ {Condition \ref{011810-24} means that the
            interaction potential $\frac12 r^2+\nu U(r)$, between the
             neighboring atoms does not become ''flat'' in some
          intervals, which would prevent the energy transfer in the chain.}}
        \begin{equation}
          \label{011810-24}
          r+\nu U'(r) =0
\end{equation}
{has only a finite number of solutions.}}
        Then,
        \begin{equation}
          \label{051705-24a}
            {\cal W}_N(\nu)=\frac{\ga}{\theta}\Big( \int_0^{\theta}p_{-N, {\rm p}}^2(s)\dd s+\int_0^{\theta}p_{N, {\rm p}}^2(s)\dd s\Big)>0.
          \end{equation}
         {In addition,
          \begin{equation}
          \label{051705-24b}
            \lim_{N\to+\infty}{\cal W}_N(\nu)=0.
          \end{equation}}
      \end{theorem}
      {The proof of this result is presented in Section \ref{sec4.4}.}

\subsection{Spatial decay  of the periodic solution as $N\to\infty$}
\label{sec:prop-peri-solut}

\begin{theorem}
  \label{thm:decay}
  There exist positive constants $C_1, A,\rho$ depending only on $\omega, \gamma$
  and independent of $N$ such that
  for $|\nu|< \nu_0\wedge C_1$ 
    \begin{equation}
         \label{exp:decay}
         \begin{split}
            &        |q_{x,p}(t;\nu)|\le A\exp\left\{-\rho|x|\right\},\quad
            t\in[0,\theta] \quad\mbox{and}\\
            &
           \int_0^{\theta}p^2_{x,p}(t;\nu)\dd t\le A\exp\left\{-\rho|x|\right\},\quad
            \,x\in\bbZ_N.
\end{split}
\end{equation}
\end{theorem}
{The proof of this result is presented in Section \ref{secC.3}.}

\subsection{The case of odd harmonic modes: loss of uniqueness}
\label{rmk3.6}

    It turns out that in the special
    case  { when both potentials $V(\cdot)$, $U(\cdot)$ are even}  and
    % and the forcing $t\mapsto{\cal F}(t)$ has only odd modes, i.e. if 
    \begin{equation}
      \label{011902-25}
      \begin{split}
      \hat{\rm
        F}_{2m}=0 \quad\mbox{for all }\quad m\in\bbZ
   %   \quad\mbox{and both potentials $V(\cdot)$, $U(\cdot)$ are even,}
     \end{split}
   \end{equation}
   the convergence of the series \eqref{010401-24z} can be extended to all
   $|\nu|< \bar\nu_0 = \frac{\bar\delta_*}{ {\frak V}}$, where
   we can define
    \begin{equation}
    \label{delta-m0}
    \bar\delta_*:=\inf\Big[\big|\big((2m-1)\om\big)^2-w^2\big|
    :\,m\in\bbZ,\,w\in {\cal I}\Big] > \delta_*.
  \end{equation}

  \begin{theorem}
  \label{main:thm2}
 % Suppose that $N=0,1,\ldots$. 
 Under hypothesis \eqref{011902-25}  
estimates \eqref{082302-24} are in force  
for $|\nu|< \bar\nu_0$, Consequently in this case  
  the series \eqref{010401-24z}
 defines a $\theta$-periodic solution of \eqref{012805-24x}.
  \end{theorem}

   {  The proof of this result is contained in Section \ref{rmk012501-25}.}

\begin{remark}
  \label{rmk011902-25}
  We stress here the fact that \autoref{main:thm2} does not claim either the uniqueness, or
   stability of the solution   given by the series \eqref{010401-24z},
   in the case $\nu_0\le |\nu| <\bar\nu_0$.
   In this interval the periodic solution  {defined by
     \eqref{010401-24z} still exists 
   but it may    be  neither stable nor unique.
   As we shall see in Section \ref{sec4.5}, the system \eqref{012805-24x} may
   admit more than one $\theta$-periodic solution.} Obviously, in that case, the
   stability in the sense claimed in Theorem \ref{thm011111-24b}
   cannot hold.
  \end{remark}

\section{The case $N=0$: a single anharmonic
  oscillator with damping}
\label{sec:case-n=0}

The case when $N=0$, i.e. a single anharmonic oscillators
with damping and
 {a one mode} periodic forcing (see  \eqref{012301-25}), is interesting in its own
right and is an example of what happens for small $N$, \cite{Dyakonov_1}.
It is instructive to look first at the harmonic case $(\nu=0)$, where the equation
can be solved explicitly. {Assuming for simplicity that ${\cal
    F}(t)=F\cos(\om t)$,
equation \eqref{012301-25} is given by
\begin{equation}
\ddot q(t)=-\omega_0^2 q(t)-2\gamma \dot q(t)+F\cos(\om t)\quad,\quad q(0)=q_0,\quad\dot q(0)=p_0 .\label{qtotal}
\end{equation}
Its  solution when  $\omega_0\not=\om$ and $\gamma=0$ is given by:
\begin{eqnarray}
  \label{qtotal2}
  q(t)=q_0 \cos(\omega_0 t)+\frac{p_0}{\omega_0}\sin(\omega_0
   t)+F\,\frac{\cos(\omega t)-\cos(\omega_0 t)}{\omega_0^2-\om^2}.\label{q01}
\end{eqnarray}
When $\om=\om_0$ (the resonance case) we interpret the solution as
 the limit $\om\to\om_0$.
 
If $\gamma>0$, then denoting $\lambda_\pm=\gamma\pm
\sqrt{\ga^2-\om_0^2}$ we can write the solution of \eqref{qtotal}
\begin{eqnarray}
&&q(t)= \alpha_0 e^{-\lambda_+ t}+\alpha_1 e^{-\lambda_- t}\nonumber\\
&&\phantom{1234}+\frac{F}{(\omega_0^2-\omega^2)^2+4\gamma^2\omega^2}\left[(\omega_0^2-\omega^2)\cos(\omega t)+2\gamma\omega\sin(\omega t) \right]\label{q0}
\end{eqnarray}
and
\[
  \begin{split}
    &
  \alpha_0=\frac{1}{2\sqrt{\ga^2-\om_0^2}}\Big[ \frac{2\ga F
    \omega^2}{(\omega_0^2-\omega^2)^2+4\gamma^2\omega^2}-p_0\Big]-\frac{\lambda_-}{2\sqrt{\ga^2-\om_0^2}}\Big[q_0- \frac{F
    (\omega_0^2-\omega^2)}{(\omega_0^2-\omega^2)^2+4\gamma^2\omega^2}\Big],\\
  &
   \alpha_1=\frac{\lambda_+}{2\sqrt{\ga^2-\om_0^2}}\Big[q_0- \frac{F
    (\omega_0^2-\omega^2)}{(\omega_0^2-\omega^2)^2+4\gamma^2\omega^2}\Big]-\frac{1}{2\sqrt{\ga^2-\om_0^2}}\Big[ \frac{2\ga F
    \omega^2}{(\omega_0^2-\omega^2)^2+4\gamma^2\omega^2}-p_0\Big].
   \end{split}
 \]
  {For  $\ga=\om_0$ $(\la_+=\la_-=\ga)$ we  take the limit  
 $\ga\to\om_0$ to get a finite value.}

 \begin{remark}
   \label{rmk011003-25}
Notice that  when  $ \gamma>0$, the system tends to a $\theta-periodic$
state for any initial condition ($\text{Re}(\lambda_\pm)>0$).  {However,
when $\gamma=0$ and $\om\not=\om_0$ only the
specific initial condition  
$$
q_0=\frac{F}{\omega_0^2-\omega^2},\quad p_0=0
$$ drives the system to such a periodic state. For other initial data
there is no  $\theta$-periodic
solution. For $\om=\om_0$ no $\theta$-periodic solution exists.}
\end{remark}}
\bigskip

Under  the anharmonic perturbation Theorems \ref{main:thm} --
\ref{thm011111-24b} hold with the following modifications:
     \begin{equation}
    \label{delta-m0}
    \begin{split}
      &  \delta_*:= \inf_{m=0,1,\ldots} \phi (m),\quad\mbox{where}\\
      &
      \phi(m):=\left\{\Big( \om_0^2  -(m\om)^2\Big)^2 +4\ga^2(m\om)^2\right\}^{1/2}.
      \end{split}
    \end{equation}
    A simple calculation shows that for $\om_0>\sqrt{2}\ga$
     \begin{equation}
    \label{032301-25}
    \delta_*= \left\{
      \begin{array}{ll}
        \min \{\phi(m_*), \phi(m_*+1)\},&\mbox{where}\\
        & \\
        m_*:=\floor*{\frac{(\om_0^2-2\ga^2)^{1/2}}{\om}}&
        \end{array}
      \right.
    \end{equation}
    and for $\om_0\le \sqrt{2}\ga$ we have
    \begin{equation}
    \label{032301-25a}
    \delta_*= \om_0^2.
  \end{equation}
 
 %  We let
%      \begin{equation}
%       \label{nu01}
%       \nu_0:=\frac{\delta_*}{\|V'\|_\infty }.
% \end{equation}

For the situation examined in Section \ref{rmk3.6} ($V$ and $U$ even and only odd forcing modes),
in the case $N=0$ the value of $\bar \delta_*$ is
     given by formula \eqref{032301-25}, if $\om_0\ge
    \sqrt{\om^2+2\ga^2}$ and  
     \begin{equation}
    \label{delta-m1}
     \bar\delta_*:=\left\{\Big( \om^2-\om_0^2  \Big)^2 +4(\ga\om)^2\right\}^{1/2},
   \end{equation}
   if  $\om_0< \sqrt{    \om^2+2\ga^2}$.
Then we define $ \bar\nu_0:=\frac{\bar\delta_*}{{\frak V} }$.
   % \begin{equation}
   %    \label{Bnu0}
   %    \bar\nu_0:=\frac{\bar\delta_*}{{\frak V} }.
   %  \end{equation}

{Finally, observe that the values $\nu_0$ (or $\bar\nu_0$) for $N=0$ depend explicitly on the friction coefficient $\gamma$. We expect that an optimal $\nu_0$ would generally depends on $\gamma$ and $N$ and, therefore, our estimation (\ref{nu0}) is just a lower bound of all of them that is indepent of $N$ and $\gamma$. }

\section{Proof of the convergence of the perturbative series.}
\label{sec:proof-conv-pert}

We prove in this section the convergence of the perturbative scheme, Theorem \ref{main:thm}
and the corresponding uniqueness result, Theorem \ref{thm-unique}.

  \subsection{Time harmonics of a periodic solution}

\label{sec:time-harm}
  
% In Section \ref{sec3}
Consider a  
$\theta$-periodic
solution to \eqref{012805-24x}, i.e. such a solution $\big(
q_{x,{\rm p}}(t;\nu)\big)$   that satisfies
\begin{equation}
\label{010810-24}
q_{x,{\rm p}}(t+\theta;\nu) =q_{x,{\rm p}}(t;\nu),\quad
t\in\bbR,
\end{equation}
Then,
\begin{equation}
  \label{011211-24}
q_{x,{\rm p}}(t;\nu)=:\sum_{m\in\bbZ}\tilde
q_x (m;\nu )e^{im\om t}.
\end{equation}
where
\begin{equation}
\label{020810-24}
\tilde q_{x}(m;\nu) =\frac{1}{\theta}\int_0^{\theta}e^{-im\om
  t}q_{x,{\rm p}}(t;\nu)\dd t,\quad
m\in\bbZ,\,x\in\bbZ_N
\end{equation}
are the Fourier coefficients  of the $\theta$-periodic solution.
They satisfy the system of equations
\begin{equation} 
\label{010907-24}
\begin{aligned}
   0&=  \big[(\om m)^2-\om_0^2+ \Delta -i\ga  \om m 
    (\delta_{x,-N}+\delta_{x,N})\big]\tilde
  q_x (m;\nu ) +\hat{\rm  F}_m \delta_{x,0}\\
  & +\nu\tilde v_{x}(m;\nu)
    , \; \quad x\in \bbZ_N,
  \end{aligned} \end{equation}
with 
\begin{equation}
  \label{030810-24}    
  \begin{split}
&\tilde v_{x}(m;\nu)= \frac{1}{\theta}\int_0^\theta e^{-im\om  t}
v_{x }(t;\nu) \dd t \quad\mbox{and}\\
&
v_{x }(t;\nu) :=-V'\big(q_{x,{\rm p}}(t;\nu)\big)+\nabla
U'\big(q_{x,{\rm p}}(t;\nu)-q_{x-1,{\rm p}}(t;\nu)\big),
\quad (m,x)\in\bbZ\times\bbZ_N.
\end{split}
\end{equation}
Denote by $\ell_{2,N}$ the space of all square integrable complex
valued sequences $\tilde {\bf f}:=\big(\tilde f_x(m)\big) _{(m,x)\in\bbZ\times\bbZ_N}$
satisfying
$\tilde f_x^\star(m)=\tilde f_x(-m)$ equipped with the Hilbert space
norm
\begin{equation}
  \label{021602-25}
\|\tilde {\bf f}\|_{\ell^2,N}:=\left(\sum_{(m,x)\in\bbZ\times\bbZ_N}|\tilde f_x(m)|^2\right)^{1/2}.
\end{equation}
Immediately, from \eqref{010907-24} and Green's function estimate \eqref{021110-24}  we conclude the following.
\begin{proposition}
  \label{prop010810-24}
  Suppose that {$F\in L^2[0,\theta]$}  and $\big( \tilde
  q_x (m;\nu )\big)_{(m,x)\in\bbZ\times\bbZ_N}\in \ell_{2,N}$ is a solution of the
  system \eqref{010907-24}. {Then,  
   ${\bf q}(t):=\big( q_{x,{\rm p}}(t;\nu)\big)
    _{(t,x)\in\bbZ\times\bbZ_N}$  given by \eqref{011211-24} belongs
  to $H^2\big([0,\theta];\bbR^{2N+1}\big)$ and
    is a periodic solution of \eqref{012805-24x}.}
  \end{proposition}

% The main goal of the present section is to formulate the result on the
% existence of a $\theta$-periodic solution of \eqref{012805-24x}.
% An essential assumption we shall be using throughout this note is that
% no integer multiplicity of the forcing frequency belongs to the
% spectrum of the harmonic chain on the lattice $\bbZ$, with no boundary
% condition, i.e.
% \begin{equation}
% \label{A}
% m\om\not\in {\cal I}:=[\om_0,\om_u]\quad\mbox{ for all }m\in\bbZ.
% \end{equation}
% Here $\om_u:=\sqrt{\om_0^2+4}$.

\subsection{The case of a harmonic chain}

Consider first the case of a harmonic chain, i.e. when $\nu=0$. 
Then  equation  \eqref{012805-24x} takes the form 
\begin{equation} 
\label{012805-24y}
\begin{split}   \ddot    q_x(t) =  \big(\Delta  -\om_0^2\big)q_x(t) 
  -\ga
   \big(\delta_{-N}(x)+
   \delta_{N}(x)\big) \dot q_x(t)  + {\cal F}(t/\theta)\delta_{x,0} ,\quad
   x\in \bbZ_N.
   \end{split}
 \end{equation}
  
 \begin{proposition}
   \label{prop010309-24}
   % Suppose that $N=0,1,2,\ldots$. Then,
   For $\nu = 0$ there exists a unique $\theta$-periodic   solution.
  % ${\bf q}_{\rm p}(t)=\big(q_{{\rm p},x}(t)\Big)_{x\in\bbZ_N}$.
   \end{proposition}
\proof
The Fourier coefficients 
% \begin{equation}
%   \label{eq:6a}
%   \tilde q_x(m)=\frac{1}{\theta}\int_0^{\theta} e^{- im\om
%   t} q_x(t)\dd t,\quad (m,x)\in\bbZ\times\mathbb Z_N
% \end{equation}
of a $\theta$-periodic  solution 
 satisfy
\begin{equation} 
\label{eq:qdynamicsbulk-av-f1}
 \Big( \om_0^2-\Delta-(m\om)^2 +i\ga m\om(\delta_{-N}+\delta_{N})\Big)  \tilde  q_x(m)= \hat{\rm F}_m\delta_{x,0}
 , \; \quad x\in\bbZ_N,\quad m\in\bbZ.
  \end{equation}

%\subsection{The case frequencies of the forcing fall outside ${\cal I}$}

\label{sec4.1}

Then,  the unique   solution of
\eqref{eq:qdynamicsbulk-av-f1} is given by the formula 
  \begin{align}
    \label{010309-24}
      \tilde  q_x(m)=\hat{\rm F}_mH_{-(m\om)^2,\ga m\om}^{(N)}(x,0)
  \end{align}
  and
  \begin{align}
    \label{012110-24}
    q_{{\rm p},x}(t)=\sum_{m\in\bbZ} \tilde  q_x(m)e^{im\om t},\quad p_{{\rm p},x}(t)
    =\sum_{m\in\bbZ} im\om\tilde  q_x(m)e^{im\om t}.
  \end{align}
 Here $H_{-(m\om)^2,\si}^{(N)}(x,y)$ is the Green's function of  $\la+\om^2_0 -i\si(\delta_{-N}
 + \delta_{N} )-\Delta$, i.e.
\begin{equation} 
\label{HN}
H_{\la,\si}^{(N)}(x,y)=\big(\la+\om_0^2-\Delta  +i\si(\delta_{ -N}
+ \delta_{ N} )\big)^{-1}\delta_y(x),
\end{equation}
where we assume that $\si,\la\in\bbR$. The computation of the Green's
function is carried out in 
   Section 
\ref{secA3}. 
 \qed

 {
 \begin{remark}
   By \eqref{021205-21b} and Proposition \ref{lm021110-24}, the solution given by
   \eqref{010309-24} is well defined even if $\gamma = 0$.
   However, without the dissipation at the boundaries  this will be just a very special solution
   without any stability properties, see Remark
   \ref{rmk011003-25}. When $\gamma>0$  this is a
   globally stable periodic solution, c.f. \eqref{qtotal} and \eqref{qtotal2}.
 \end{remark}
 }

\subsection{Proof of Theorem \ref{main:thm}}

\label{sec3.3.1}
    
Consider the  Fourier coefficients  of
the approximating scheme \eqref{eq:flip-lZa}  
\begin{equation}
  \label{eq:6aZ}
  \begin{split}
\tilde q_x^{(L)}(m;\nu )=\frac{1}{\theta}\int_0^\theta e^{-im\om  t}  q_{x,{\rm p}}^{(L)}(t;\nu)dt,\quad m \in\mathbb Z_N.
\end{split}
\end{equation}
Coefficients $\tilde q_x^{(0)}(m )$   satisfy  \eqref{010907-24} for $\nu = 0$,
% \begin{equation} 
% \label{011005-21-0}
%   0 =  \left[\Delta  
%   +(m\om)^2-\om_0^2 -i\ga m\om(
%     \delta_{-N}(x)+
%     \delta_N(x))\right]  \tilde q_x^{(0)}(m ) 
%   +  \;      \hat{\rm  F}_m  \delta_{x,0}
%     , \; \quad x\in\bbZ_N
%     \end{equation}
and for $L=1,2,\ldots$  
\begin{equation} 
\label{011005-21-0l}
\begin{aligned}
  0 &=  \left[\Delta  
  +(m\om)^2-\om_0^2 -i\ga m\om(
    \delta_{-N}(x)+
    \delta_N(x))\right]  \tilde q_x^{(L)}(m ) 
 -  \;     \tilde v_{x,L-1}(m;\nu)
    , \; \quad x\in\bbZ_N,
  \end{aligned} \end{equation}
with (see \eqref{vxL}) 
\begin{equation}
  \label{022810-24}
 \tilde v_{x,L-1}(m;\nu)= \frac{1}{\theta}\int_0^\theta e^{-im\om  t}
v_{x,L-1}(t;\nu) \dd t.
\end{equation}

 {We have, see \eqref{HN},}
\begin{align}
    \label{012302-24}
    \tilde  q_x^{(0)}(m ;\nu)=\hat{\rm  F}_m  H_{-(m\om)^2, \ga m\om}(x,0)
\end{align}
and
\begin{align}
    \label{012302-24a}
    &  \tilde  q_x^{(L)}(m ;\nu)= \sum_{y\in\bbZ_N}H_{-(m\om)^2, \ga m\om}(x,y)\tilde v_{y,L-1}(m) ,\quad
  x\in\bbZ_N,\,\ell=1,2,\ldots. 
  \end{align}

Multiplying both sides of \eqref{011005-21-0l}  by $\Big( \tilde
q_x^{(L)}(m )\Big)^\star$ and summing   over  $x$   we get
\begin{align}
  \label{062410-24}
&\sum_{x\in\bbZ_N} \Big( \om_0^2 -\Delta -(m\om)^2\Big)\tilde q_x^{(L)}(m )\Big( \tilde q_x^{(L)}(m )\Big)^\star =  
   -i\ga(m\om)
   \tilde   q_{-N}^{(L)}(m) \Big( \tilde q_{-N}^{(L)}(m
    )\Big)^\star \\
 &-i\ga (m\om)
    \tilde   q_{N}^{(0)}(m) \Big( \tilde q_{N}^{(0)}(m
    )\Big)^\star 
  +  \;    \sum_{x\in\bbZ_N}   \tilde v_{x,L-1}(m ;\nu) \Big( \tilde q_x^{(L)}(m )\Big) ^\star ,\notag
\end{align}
for each $m\in\bbZ$ and $L=1,2,\ldots$.
{Observe that
  \begin{equation}
    \label{eq:7}
 0<   -\sum_{x\in\bbZ_N}  {\big(\Delta\tilde q_x^{(L)}(m )\big) \tilde q_x^{(L)}(m )^\star}=
 \sum_{x\in\bbZ_N}\Big|\nabla\tilde q_x^{(L)}(m )\Big|^2 \le
 4\sum_{x\in\bbZ_N}\big(\tilde q_x^{(L)}(m )\big)^2.
  \end{equation}
Consequently for the real part of \eqref{062410-24} we have
  \begin{equation}
    \begin{split}
    \sum_{x\in\bbZ_N} \Big( \om_0^2 -(m\om)^2\Big) |\tilde q_x^{(L)}(m )|^2
    \le {\rm Re}  \sum_{x\in\bbZ_N}   \tilde v_{x,L-1}(m ;\nu) \Big( \tilde q_x^{(L)}(m )\Big) ^\star\\
    \le \sum_{x\in\bbZ_N}  \left|\tilde v_{x,L-1}(m ;\nu)\right| \Big| \tilde q_x^{(L)}(m )\Big|.
  \end{split}
  \label{eq:5}
\end{equation}
}

Since
$
\tilde q_x^{(L)}(-m )=\Big( \tilde q_x^{(L)}(m )\Big)^\star
$
we have
$
|\tilde q_x^{(L)}(-m )|^2=|\tilde q_x^{(L)}(m )|^2.
$
Suppose that $0<|m|\om<\om_0$. Then
\begin{align}\label{eq:left}
\sum_{0<(m\om)^2<\om_0^2}\sum_{x\in\bbZ_N}
 \Big( \om_0^2  
                 -(m\om)^2\Big) |\tilde q_x^{(L)}(m )|^2 \le  
  \;   { \sum_{0<(m\om)^2<\om_0^2}\sum_{x\in\bbZ_N}  \left|\tilde v_{x,L-1}(m ;\nu)\right|
  \Big| \tilde q_x^{(L)}(m )\Big|.}
\end{align}

In the case when  $|m|\om>\sqrt{\om_0^2+4}$ we argue as follows.
Using inequality \eqref{eq:7} we have 
\begin{align}
  \label{011802-25}
  &
    \sum_{(m\om)^2>\om_0^2+4}\sum_{x\in\bbZ_N} \Big( 
                 (m\om)^2 -\om_0^2-4 \Big)|\tilde q_x^{(L)}(m
    )|^2 \notag\\
  &
    \le \sum_{(m\om)^2>\om_0^2+4}\sum_{x\in\bbZ_N} \Big[\Big( 
                 (m\om)^2 -\om_0^2 \Big)|\tilde q_x^{(L)}(m
    )|^2-\Big|\nabla\tilde q_x^{(L)}(m ) \Big|^2\Big]\\
   &
    =\sum_{(m\om)^2>\om_0^2+4}\sum_{x\in\bbZ_N} \underbrace{\Big( 
                 (m\om)^2 +\Delta-\om_0^2 -i\ga m(\delta_{-N}(x)+\delta_{x,N})\Big)\tilde q_x^{(L)}(m
    )}_{=-\tilde v_{x,L-1}(m ;\nu) }\big(\tilde q_x^{(L)}(m ) \big)^\star \notag\\
  &
    =  
   \;    -\sum_{(m\om)^2>\om_0^2+4}\sum_{x\in\bbZ_N}  \tilde v_{x,L-1}(m ;\nu) \Big( \tilde q_x^{(L)}(m )\Big) ^\star. \notag
\end{align}
{Hence we can now sum over all $m$ and by \eqref{eq:left}, \eqref{011802-25}, the definition of
$\delta_*$ \eqref{delta-m} and the Cauchy-Schwarz inequality we obtain}
  \begin{equation}
    \label{092302-24}
    \begin{split}
    \delta_*\|\tilde
  q^{(L)}(\cdot;\nu)\|_{\ell^2,N}^2
  \le
 \left( \|\tilde v_{L-1}(\cdot
   ;\nu)\|_{\ell^2,N}\right)^{1/2} \left(  \|\tilde
   q^{(L)}(\cdot;\nu)\|_{\ell^2,N}^2 \right)^{1/2} .
 \end{split}
\end{equation}
Then, by the Plancherel identity,
  \begin{align*}
 & \|\!|q^{(L)}_{\rm p}(\cdot;\nu)\|\!|^2=\|\tilde
   q^{(L)}(\cdot;\nu)\|_{\ell^2,N}^2 
      \le  \frac{1}{\delta_*} \|\tilde v_{L-1}(\cdot ;\nu)\|_{\ell^2,N}^2\\
    &
      =  \frac{1}{\delta_*\theta}\sum_{x\in\bbZ_N}\int_0^\theta [v_{x,L-1}(t ;\nu)]^2\dd  t .
  \end{align*}

% \begin{equation*}
%   \begin{split}
% &v_{x,L-1}(t;\nu)=\frac{1}{\nu^{L-1}}\Bigg[V'\big(Q_{x,{\rm
%     p}}^{(L-1)}(t;\nu)\big)-V'\big(Q_{x,{\rm p}}
% ^{(L-2)}(t;\nu)\big)\Bigg] \\
% &
% -\frac{1}{\nu^{L-1}}\Bigg[\nabla U'\big(Q_{x,{\rm
%     p}}^{(L-1)}(t;\nu)-Q_{x-1,{\rm
%     p}}^{(L-1)}(t;\nu)\big)-\nabla U'\big(Q_{x,{\rm
%     p}}^{(L-2)}(t;\nu)-Q_{x-1,{\rm
%     p}}^{(L-2)}(t;\nu)\big) \Bigg] 
% \end{split}
% \end{equation*}

  Using \eqref{vxL} and \eqref{Vp1z} we conclude that
  \begin{align}
    \label{110301-25}
& | v_{x,L-1}(t;\nu)|\le \frac{1}{|\nu|^{L-1}}\Big[\|V''\|_{\infty}\big|Q_{x,{\rm p}}^{(L-1)}(t;\nu) -Q_{x,{\rm p}}^{(L-2)}(t;\nu)\big|\notag\\
  &
    +\|U''\|_{\infty}\Big(\big|\nabla^\star \big(Q_{x,{\rm p}}^{(L-1)}(t;\nu) - Q_{x,{\rm p}}^{(L-2)}(t;\nu)\big)\big|+\big|\nabla \big(Q_{x,{\rm p}}^{(L-1)}(t;\nu) - Q_{x,{\rm p}}^{(L-2)}(t;\nu)\big)\big|\Big) \notag\\
  &
    =\|V''\|_{\infty}
  |q^{(L-1)}_{x,{\rm p}}(t;\nu)|+\|U''\|_{\infty}\Big(|\nabla^\star q^{(L-1)}_{x,{\rm
    p}}(t;\nu)|+|\nabla  q^{(L-1)}_{x,{\rm p}}(t;\nu)|\Big)\\
  &
    \le (\|V''\|_{\infty}+\|U''\|_{\infty})
  |q^{(L-1)}_{x,{\rm p}}(t;\nu)|+\|U''\|_{\infty}\Big(| q^{(L-1)}_{x-1,{\rm
    p}}(t;\nu)|+| q^{(L-1)}_{x+1,{\rm p}}(t;\nu)|\Big). \notag
\end{align}
Hence,
\begin{align}
  \label{102410-24a}
& [ v_{x,L-1}(t;\nu)]^2 
    \le (\|V''\|_{\infty}+\|U''\|_{\infty})^2
                 [q^{(L-1)}_{x,{\rm p}}(t;\nu)]^2\notag\\
  &
    +(\|V''\|_{\infty}+\|U''\|_{\infty})\|U''\|_{\infty}\Big( 2[q^{(L-1)}_{x,{\rm p}}(t;\nu)]^2+ [q^{(L-1)}_{x-1,{\rm p}}(t;\nu)]^2+ [q^{(L-1)}_{x+1,{\rm p}}(t;\nu)]^2\Big)\\
  &
    +2 \|U''\|_{\infty}\Big([ q^{(L-1)}_{x-1,{\rm
    p}}(t;\nu)]^2+[ q^{(L-1)}_{x+1,{\rm p}}(t;\nu)]^2\Big).\notag
\end{align}
As a result, see \eqref{delta-m},
    \begin{align*}
     & \|\!|q^{(L)}_{\rm p}(\cdot;\nu)\|\!|^2_N
      \le  \frac{{\frak V}^2 }{ \delta_*^{2}\theta}    \sum_{x\in\bbZ_N}\int_0^\theta
       [q^{(L-1)}_{x,{\rm p}}(t;\nu)]^2 \dd t 
   \end{align*}
   and estimate \eqref{082302-24} follows for $N=1,2,\ldots$.
The remaining conclusions of the theorem follow from the estimate.\qed

 \subsection{Regularity of the periodic solution}

 From equality \eqref{011005-21-0l} we conclude that there exists a
 constant $C>0$ such that
\begin{equation}
  \label{072410-24a}
  \begin{split}
    & (m\om)^2 \|\tilde
  q^{(L)}(m;\nu)\|_{N}
  \le
C\big(\| \tilde
  q^{(L)}(m;\nu)\|_{N}+\|\tilde v_{L-1}(m
  ;\nu)\|_{N}\big),\quad m\in\bbZ.
\end{split}
\end{equation}
From \eqref{102410-24a} we obtain in turn that
\begin{equation}
  \label{041003-25}
  \|\!| 
 v_{L}(\cdot;\nu)\|\!| 
 \le {\frak V} \|\!| 
 q^{(L)}_{\rm p}(\cdot;\nu)\|\!|,\quad L=1,2,\ldots.
 \end{equation}
Combining  \eqref{072410-24a}, \eqref{041003-25}  and
\eqref{082302-24} we conclude that there exist  constants $C_1,C_2>0$ such
that
\begin{equation}
  \label{072410-24b}
  \begin{split}
    & \sum_{m\in\bbZ}[1+(m\om)^2]^2 \|\tilde
  q^{(L)}(m;\nu)\|_{N}^2
  \le  C_1\big(\|\!|  
  q^{(L)}_{\rm p}(\cdot;\nu)\|\!|^2+\|\!| 
  v_{L}(\cdot;\nu)\|\!|^2 \big)\\
  &\le  C_2 \|\!|  
  q^{(L)}_{\rm p}(\cdot;\nu)\|\!|^2\le \frac{C_2}{\nu_0^{2L}},\quad L=1,2,\ldots.
  \end{split}
\end{equation}

\begin{proposition}
  \label{prop012410-24}
  Under the assumptions of Theorem \ref{main:thm} there exists a
  constant $C>0$ such that
  \begin{equation}
    \label{021003-25}
    \begin{split}
      &\sup_{t\in\bbR}\|q^{(L)}_{\rm p}(t;\nu)\|_N\le
      \frac{C}{\nu_0^L}\quad\mbox{and}\\
     &\sup_{t\in\bbR}\|\dot q^{(L)}_{\rm p}(t;\nu)\|_N\le
     \frac{C}{\nu_0^L} ,\quad L=1,2,\ldots.
     \end{split}
   \end{equation}
  In consequence,  there
 exists $C>0$
 \begin{equation}
   \label{112410-24}
  \sup_{t\in\bbR} \Big(\|Q^{(L)}_{\rm p} (t;\nu)\|_{N}+\|\dot Q^{(L)}_{\rm p} (t;\nu)\|_{N}\Big)\le C,\quad L=0,1,2,\ldots.
\end{equation}
Furthermore 
 $Q^{(L)}_{\rm p} (\cdot;\nu)\in
 C^2 \Big([0,\theta];\bbR^{2N+1}\big)$
and
the series given by \eqref{010401-24z} converges uniformly in $t$ to a $\theta$-periodic
function $\Big( q_{x,{\rm p}}(t;\nu)\Big)$ whose Fourier coefficients $\Big(\tilde q_x(m;\nu )\Big)$
satisfy the system \eqref{010907-24}.
  \end{proposition}
  \proof  From  bound    \eqref{072410-24b}
and the
Cauchy-Schwarz inequality  we have
  \begin{align*}
    &\sup_{t\in\bbR}\|q^{(L)}_{\rm p}(t;\nu)\|_N\le \sum_{m\in\bbZ}\|\tilde
      q^{(L)}(m;\nu)\|_{N}\\
    &
      \le  \left(\sum_{m\in\bbZ}[1+(m\om)^2]\|\tilde
      q^{(L)}(m;\nu)\|_{N}^2\right)^{1/2}
      \left(\sum_{m\in\bbZ}[1+(m\om)^2]^{-1} \right)^{1/2} \le 
     \frac{C}{\nu_0^L} 
  \end{align*}
  for some constant $C>0$. The proof of the second estimate in
  \eqref{021003-25} is similar. These estimates  obviously imply   the uniform convergence of the series
  in \eqref{010401-24z} and bound \eqref{112410-24}.

 {From \eqref{072410-24b} and its definition \eqref{010905-24} we conclude also 
  that $Q^{(L)}_{\rm p} (\cdot;\nu)$ belongs to
$ H^2 \Big([0,\theta];\bbR^{2N+1}\big)$ and satisfies
 \eqref{012805-24LL}. In fact $Q^{(L)}_{\rm p} (\cdot;\nu)\in
 C^2 \Big([0,\theta];\bbR^{2N+1}\big)$. }
  \qed

% Let ${\bf  q}_{{\rm
%     p}}(t;\nu)=\Big( q_{x,{\rm p}}(t;\nu)\Big)$ and
% ${\bf  p}_{{\rm
%     p}}(t;\nu)=\Big( \dot q_{x,{\rm p}}(t;\nu)\Big)$.
% The Hamiltonian of the chain whose dynamics is described by
% \eqref{012805-24x} equals
% \begin{equation}
%   \label{Hal}
%   \begin{split}
%  &   {\cal H}({\bf q},{\bf p})=\sum_{x\in\bbZ_N}e_x({\bf q},{\bf p};\nu),\quad\mbox{where},\\
% &
% e_x({\bf q},{\bf p};\nu):=  \frac{p_x^2}2 +
% \frac12 (q_{x}-q_{x-1})^2 +\frac{\om_0^2 q_x^2}{2}+\nu\Big( V(q_x)+
% U(q_x-q_{x-1})\Big),\quad ({\bf q},{\bf p})\in \Omega_N.
% \end{split}
% \end{equation}
 
  From the proof of Theorem \ref{main:thm} we conclude also the following.
\begin{corollary}
  \label{cor010810-24}
 Under the assumptions of Theorem \ref{main:thm} for any
 $|\nu|<\nu_0$ there exists a constant $C>0$ independent of $N$ and $\ga>0$  {such
 that the Hamiltonian ${\cal H}_N(\cdot,\cdot)$ defined in
 \eqref{010503-25} satisfies}
 \begin{equation}
   \label{060810-24}
  \frac{1}{\theta}\int_0^\theta{\cal H}_N\big({\bf  q}_{{\rm
      p}}(t;\nu), {\bf  p}_{{\rm p}}(t;\nu)\big)\dd t\le C.
   \end{equation}
  \end{corollary}

%   \medskip

%   \begin{corollary}
  
%     \end{corollary}

 %\textcolor{red}{\bf DOTAD}

 \subsection{Uniqueness: Proof of Theorem \ref{thm-unique}}

% Concerning the uniqueness of the periodic solution described in
% Theorem \ref{main:thm}   we have  the following result.
% \begin{theorem}
%   \label{unique}
%   Suppose that $\nu_0$ is given by formula \eqref{nu0}, with
%   $\delta_*$ defined in \eqref{delta-m} for $N=1,2,\ldots$ and
%   \eqref{032301-25}, or \eqref{032301-25a} for $N=0$. Then, for $|\nu|< \nu_0$ the system
%   \eqref{012805-24x} has a unique $\theta$-periodic solution in $L^2\big([0,\theta];\bbR^{2N+1}\big)$.
%   \end{theorem}
  \proof
  Consider the mapping ${\cal T}$ of
  $L^2\big([0,\theta]; \bbR^{2N+1}\big)$ into itself, assigning to
  each $f(t)=\big(f_x(t)\big)$ belonging to 
$ L^2\big([0,\theta]; \bbR^{2N+1}\big)$ an element
  $$
 ({\cal T}f)_x(t)=\sum_{m\in\bbZ}\tilde{Tf}_x(m) e^{im\om t},\quad
 x\in\bbZ_N.
  $$ given by
  \begin{equation}
    \label{020907-24}
    \begin{aligned}
 & \tilde{Tf}_x(m) =   H^{(N)}_{-(m\om)^2,\ga m\om}(x,0)\hat {\rm F}_m
  \\
  &
  \qquad  \qquad  \qquad +      \nu \sum_{y\in\bbZ_N} H^{(N)}_{-(m\om)^2,\ga m\om}(x,y)\tilde v_{y}(m;f)
    , \; \quad (m,x)\in\bbZ\times \bbZ_N. 
  \end{aligned}
\end{equation}
Here
\begin{equation}
  \label{012811-24}
  f_x(t)=\sum_{m\in\bbZ}\tilde{f}_x(m) e^{im\om t},\quad x\in\bbZ_N
  \end{equation}
  and
  $$
\tilde v_{y}(m;f)=-\frac{1}{\theta}\int_0^{\theta}\Big[V'(f_x(t))-\nabla U\big(f_x(t)-f_{x-1}(t)\big)\Big]\dd t.
  $$
  We have
   $
 \Big(\tilde{Tf}_x(m) \Big)^\star=\tilde{Tf}_x(-m) .
 $

 Suppose that  $f^{(1)}, f^{(2)}\in
  L^2\big([0,\theta];\bbR^{2N+1}\big)$. Then,
$
\delta\tilde{Tf} _x(m) :=\tilde{Tf}^{(2)}_x(m)- \tilde{Tf}^{(1)}_x(m) 
$  satsify
  \begin{equation}
    \label{030907-24}
  \begin{aligned}
  0 &=  \Big(\Delta+ (m\om)^2-\om_0^2 -i\ga m\om(
    \delta_{-N} +\delta_{N})\Big)\Big(\delta\tilde{Tf} _x(m) \Big)
  +\\
  & 
  +  \;     \nu\Big(\tilde v_{x}(m;f^{(2)})-  \tilde v_{x}(m;f^{(1)})\Big)
    , \; \quad (m,x)\in\bbZ\times \bbZ_N.
  \end{aligned}
\end{equation}
Multiplying both sides by $\Big(\delta\tilde{Tf}^{(j)}_x(m)\Big)^\star
$ and summing over $x$ we get
\begin{align*}
  & \sum_{x\in\bbZ_N}\Big[\big(\om_0^2-(m\om)^2 \big)|\delta\tilde{Tf}
    _x(m) |^2+|\nabla\delta\tilde{Tf}
    _x(m) |^2\Big]  +i\ga m\om\Big( |\delta\tilde{Tf}_{-N}(m)|^2
       +|\delta\tilde{Tf}_{N}(m)|^2\Big)\\
       &
  =\nu \sum_{x\in\bbZ_N}\Big(\tilde v_{x}(m;f^{(2)})-\tilde
         v_{x}(m;f^{(1)})\Big)  \Big(\delta\tilde{Tf}_x(m)\Big)^\star 
   .
\end{align*}
Using the assumption \eqref{A} and applying the Cauchy-Schwarz
inequality on the right hand side
  we conclude that   
\begin{align*}
  & \delta_*\sum_{m\in\bbZ}\sum_{x\in\bbZ_N}\Big|\delta\tilde{Tf}_x(m) \Big|^2\\
  &
    \le   |\nu| \left(\sum_{m\in\bbZ}\sum_{x\in\bbZ_N}\Big|\tilde v_{x}(m;f^{(2)})-\tilde
         v_{x}(m;f^{(1)})\Big|^2 \right)^{1/2}\left(\sum_{m\in\bbZ}\sum_{x\in\bbZ_N}\Big|\delta\tilde{Tf}_x(m)\Big|^2 \right)^{1/2}
   .
\end{align*}
In consequence
\begin{align*}
   |\!\|Tf^{(2)}-Tf^{(1)}|\!\|_N\le
  \frac{|\nu|}{\nu_0}|\!\|f^{(2)}-f^{(1)}|\!\|_N  .
\end{align*}
By the Banach contraction principle there exists a unique solution of
the equation ${Tf} =f$ in  $ 
 L^2\big([0,\theta];\bbR^{2N+1}\big)$. Then $f(t)$ given by
 \eqref{012811-24} is the unique $\theta$-periodic solution of
 \eqref{012805-24x}. Hence
the conclusion of the theorem follows.
\qed

 \medskip

\section{Proof of Theorem \ref{main:thm2}}
\label{rmk012501-25}

Using the argument from the proof of Theorem \ref{main:thm} one can
infer the conclusions of Theorem \ref{main:thm2}, provided 
  it can be shown that
\begin{equation}
  \label{010102-25a}
  \tilde  q_x^{(L)}(2m ;\nu)=0,\quad x\in\bbZ_N,\,L,m=0,1,2,\ldots.
\end{equation}
This would follow, if we can prove that
\begin{equation}
  \label{010102-25}
\tilde v_{x,L}(2m;\nu)=0,\quad x\in\bbZ_N,\,L,m=0,1,2,\ldots,
\end{equation}
see \eqref{022810-24}. 
To see \eqref{010102-25} (and \eqref{010102-25a}) assume for simplicity
that $U\equiv0$. The argument in the general case is similar.  Since $V'(\cdot)$ is odd and real valued, its Fourier
transform $\widehat{ V'}(\xi)$ is also an odd function.  To further
simplify the argument we suppose that its support is compact. The
general case can be treated by an approximation argument.
Thanks to \eqref{010309-24} and the fact that
$\hat F_{2m}=0$ for
all $m$,  we conclude that \eqref{010102-25a} holds
for $L=0$.
 Suppose now that \eqref{010102-25}  is true for some non-negative
 integer $L-1$. Since 
\begin{align*}
&q^{(L-1)}_{x}(t)= \sum_{m=1}^{+\infty}\tilde q^{(L-1)}_{x}(2m-1)e^{i(2m-1)\om t}+
                 \sum_{m=1}^{+\infty}\Big(\tilde q^{(L-1)}_{x}(2m-1)\Big)^\star e^{-i(2m-1)\om t}
\end{align*}
we have
\begin{align}
  \label{020102-25}
&\tilde  v_{x,L}(2m)
    =
  \frac{1}{2\pi \theta}\int_{\bbR}\hat{V'}(\xi)  \dd\xi
  \int_0^\theta e^{- 2i\om m t}\exp\left\{i\xi  
                 q^{(L-1)}_{x}(t)\right\}\dd t\notag\\
  &
    =\lim_{M\to+\infty}\frac{\theta}{2\pi i}\int_{\bbR}\hat{V'}(\xi)  \dd\xi  \int_0^\theta e^{- 2i\om m t}\exp\left\{i\sum_{m'=1}^{M}\xi \tilde
    q _{x}^{(L-1)}\big(2m'-1\big)e^{i(2m'-1)\om t}\right\}\notag\\
  &
    \times\exp\left\{i\sum_{m''=1}^{M}\xi 
   {  (\tilde q^{(L-1)}_{x}(2m''-1))^\star }e^{-i(2m''-1)\om t}\right\}\dd t.
\end{align}
\marginpar{ {Corr.!}}
Denote by $C(1)=[z\in\mathbb C:\,|z|=1]$, a contour oriented counter-clockwise.
Using  contour integration we can rewrite the utmost right hand
side of \eqref{020102-25} as 
\marginpar{ {Corr.!}}
\begin{align*}
 &
    \lim_{M,N\to+\infty}\int_{\bbR}\hat{V'}(\xi) \hat
   v^{(L)}_{M,N}(2m;\xi)  \dd\xi ,\quad\mbox{where}\\
  &
   \hat
   v^{(L)}_{M,N}(2m;\xi) := \frac{1}{2\pi i}\int_{C(1)}\prod_{m'=1}^{M}\sum_{n_{m'}=0}^{N}\frac{\big(i\xi \tilde
    q _{x}^{(L-1)} (2m'-1)z^{2m'-1}\big)^{n_{m'}}}{n_{m'}!}\\
  &
    \times\prod_{m''=1}^{M}\sum_{n'_{m''}=0}^{N}\frac{(i\xi { (\tilde
    q _{x}^{(L-1)}(2m''-1))^\star}
    )^{n'_{m''}}}{(z^{2m''-1})^{n'_{m''}}(n'_{m''})!}\frac{\dd z}{z^{2m+1}}
\end{align*}
Performing the contour integration we get
\begin{align*}
  &
    \hat
   v^{(L)}_{M,N}(2m;\xi)  =\sum_{n_1=0}^{N}\ldots
    \sum_{n_M=0}^{N}\sum_{n_1'=0}^{N}\ldots \sum_{n_M'=0}^{N}
    \big(i\xi\big)^{n_1+\ldots+n_M+n_1'+\ldots+n_M'}\\
  &
    \times \delta\Big(n_1+3n_2+\ldots+(2M-1)n_M-2m-\big(n_1'+3n_2'+\ldots+(2M-1)n_M'\big)\Big)\\
  &
    \times\prod_{m'=1}^{M}\frac{\big(\tilde
    q _{x}^{(L-1)} (2m'-1)\big)^{n_{m'}}}{n_{m'}!} \prod_{m''=1}^{M}\frac{\big( {( \tilde
    q _{x}^{(L-1)}(2m''-1))^\star}
    )^{n'_{m''}}} {(n'_{m''})!} .
\end{align*}
%\marginpar{ {Corr.!}}
Here $\delta(k)=0$, if $k\not=0$ and $1$ if $k=0$. The expression
above is non-zero only if
$$
n_1'+n_2'+\ldots+n_M'+n_1+n_2+\ldots+n_M\quad\mbox{is even}.
$$
Since $\xi\mapsto \hat{V'}(\xi)$ is odd and $\xi\mapsto\hat
   v^{(L)}_{M,N}(2m;\xi)$ is even  we conclude therefore that
$$
\int_{\bbR}\hat{V'}(\xi) \hat
   v^{(L)}_{M,N}(2m;\xi)  \dd\xi=0
   $$
and \eqref{010102-25} follows by an induction argument.\qed

\subsection{Remark about non-uniqueness}

\label{sec4.5}
Suppose that the interaction potential $U\equiv0$ and $V$ is even. If $\nu$ is as in
the statement of Theorem \ref{thm-unique}, then the derivative of the pinning potential
$q\mapsto W(q;\nu):=\frac{(\om_0 q)^2}{2}+\nu V(q)$ is strictly
increasing. Therefore, the potential attains its unique minimum at
$q=0$. Our result states that the periodic solution of
\eqref{012805-24x} constructed  in Theorem \ref{main:thm}  is unique.

However, as it has been observed in Remark \ref{rmk3.6}, the
perturbative scheme may work for a larger range of the anharmonicity
parameter $\nu$ than considered in Theorem \ref{thm-unique}. In the
latter case there could exist some periodic solutions of
\eqref{012805-24x} other than defined by \eqref{010401-24z}. Consider for example the forcing given by
\begin{equation}
  \label{F-cF}
{\cal F}\Big(\frac{t}{\theta}\Big)=2{\rm F}\cos(\om t).
\end{equation}
Assume that $V(\cdot)$ is of $C^2$ class of regularity. 
According to  Remark \ref{rmk3.6}, by adjusting $\om$ to be
sufficiently large, we can   make $\bar\nu_0$ - the range of the
convergence of the perturbative scheme - as large as we wish.
Then for some $|\nu|<\bar\nu_0$ the potential $W(q;\nu)$ could admit
some other local minimum, say at $\bar q\not=0$. Then
$W'(\bar q;\nu)=\om_0^2\bar q+\nu V'(\bar q)=0$.
Assume furthermore
that $W''(\bar q;\nu)=\om_0^2+\nu V''(\bar q)>0$. 
Then, obviously
\begin{equation}
  \label{barq}
  \bar q_x(t;0)\equiv \bar q, \quad x\in\bbZ_N
\end{equation}
solves equation
\eqref{012805-24x} with ${\cal F}(\cdot)\equiv 0$ (corresponding to ${\rm F}=0$). 
It can be seen, see Section \ref{appF} of the Appendix, that there will be  $\theta=2\pi/\om$-periodic solutions $\big(\bar q_x(t;F)\big)_{x\in
  \bbZ_N}$  of
\eqref{012805-24x} that are close to $\big(\bar q_x(t;0)\big)_{x\in
  \bbZ_N}$ for small values of $F$.

\section{Proof of Theorem \ref{prop011705-24a}}
\label{sec4.4}

        \subsection{Proof of  positivity of
          the work functional}

{The fact that ${\cal W}_N(\nu)\ge 0$ is a consequence of the
  identity \eqref{061705-24}.}
    The only remaining part is the proof of strict positivity of
        the work functional. 
        Suppose that  ${\cal W}_N(\nu)=0$. Then
        $p_{-N, {\rm p}}(t)=p_{N, {\rm p}}(t)\equiv0$, $t\ge0$. This
        implies  that $q_{-N, {\rm p}}(t) \equiv q_{-N, {\rm p}}$,  $q_{N, {\rm p}}(t) \equiv
        q_{N, {\rm p}}$, for all $t\ge 0$, and some constants $q_{-N, {\rm p}}$ and $q_{N, {\rm p}}$.
         {Then $q_{-N+1, {\rm p}}(t)$ must solve the equation
        \begin{equation*}
          0 = q_{-N+1, {\rm p}}(t) - q_{-N, {\rm p}} - \omega_0^2 q_{-N, {\rm p}}
          - \nu \left[V'(q_{-N, {\rm p}}) - U'(q_{-N+1, {\rm p}}(t) - q_{-N, {\rm p}})\right].
        \end{equation*}
        % i.e.
        % \begin{equation*}
        %   \mathcal U(q_{-N+1, {\rm p}}(t) - q_{-N, {\rm p}}) = \omega_0^2 q_{-N, {\rm p}}
        %   + \nu V'(q_{N, {\rm p}}),
        % \end{equation*}
        that implies that $q_{-N+1, {\rm p}}(t)$
        is constant in $t$ and   ${\dot p}_{-N+1, {\rm p}}(t)= 0$.
       Consequently
         \begin{equation*}
           \begin{split}
           0 = {\dot p}_{-N+1, {\rm p}}(t)
           = q_{-N+2, {\rm p}}(t) + q_{-N, {\rm p}} - (1 + \omega_0^2) q_{-N+1, {\rm p}}\\
           - \nu \left[V'(q_{-N+1, {\rm p}}) + U'(q_{-N+2, {\rm p}}(t) - q_{-N+1, {\rm p}})
             - U'(q_{-N+1, {\rm p}} - q_{-N, {\rm p}})\right].
         \end{split}
       \end{equation*}
       that implies $q_{-N+2, {\rm p}}(t)$
       is constant in $t$ and   ${\dot p}_{-N+2, {\rm p}}(t)= 0$.
       Iterating we have
         \begin{equation*}
           q_{-N+j, {\rm p}}(t) \equiv q_{-N+j, {\rm p}}, \qquad {\dot p}_{-N+j, {\rm p}}(t)= 0,
           \quad j=0,\ldots,N.
         \end{equation*}
Repeating the same argument starting from the other side we obtain
            \begin{equation*}
           q_{N-j, {\rm p}}(t) \equiv q_{N-j, {\rm p}}, \qquad {\dot p}_{N-j, {\rm p}}(t)= 0,
           \quad j=0,\ldots,N.
         \end{equation*}
This implies for the $0$-site
         \begin{equation*}
           \begin{split}
            0=  {\dot p}_{0, {\rm p}}(t) = &( q_{1, {\rm p}} - q_{-1, {\rm p}} - 2 q_{0, {\rm p}})
             - \omega_0^2 q_{0, {\rm p}}\\
             &- \nu \left[V'(q_{0, {\rm p}}) + U'(q_{1, {\rm p}} - q_{0, {\rm p}}(t)) -
            U'(q_{0, {\rm p}} - q_{-1, {\rm p}}) \right]
             + {\cal F}(t/\theta)
          \end{split}
         \end{equation*}
and we  conclude that ${\cal F}(t/\theta)\equiv {\rm const}$, which contradicts \eqref{forcing-g}.}
        \qed

  \medskip

  \subsection{Proof of  (\ref{051705-24b})}

 { To conclude \eqref{051705-24b} it suffices to show that
  \begin{equation}
    \label{011404-25}
 \lim_{N\to+\infty}  \frac{1}{\theta}\int_0^{\theta}\big[p_{\pm N,
  {\rm p}}( t;\nu)\big]^2\dd t =0.
\end{equation}
From Proposition \ref{prop012410-24} we have
\begin{align*}
  p_{\pm N, {\rm p}}(t;\nu)=\sum_{L=0}^{+\infty}\nu^L \dot q^{(L)}_{\pm N,
  {\rm p}}(t;\nu),
\end{align*}
where the series converges  uniformly in $t$, as $N\to+\infty$, for $|\nu|<\nu_0$. Using this and
estimate \eqref{072410-24b}, to conclude
\eqref{011404-25} it suffices to prove that 
\begin{equation}
    \label{011404-25a}
 \lim_{N\to+\infty}  \frac{1}{\theta}\int_0^{\theta}\big[q^{(L)}_{\pm N,
  {\rm p}}( t;\nu)\big]^2\dd t=  \lim_{N\to+\infty} \sum_{m\in\bbZ} |\tilde q_{\pm N}^{(L)}(m
    )|^2 =0
\end{equation}
for each $L=0,1,\ldots$. The latter is a consequence of formulas
\eqref{012302-24} and \eqref{012302-24a} and  estimate 
  \eqref{071110-24}.}

\section{Global stability of the periodic solution}

\label{sec4}

In the present section we assume that  
$N$ is finite and 
$
{\bf q}\big(t;\nu, {\bf q}, {\bf p}\big)=\Big(q_x\big(t;\nu, {\bf q}, {\bf p}\big)\Big),
$ is the solution of \eqref{012805-24x}  with the initial condition {at $s$}
\begin{equation}
  \label{011109-24}
  {q_x\big(s;\nu, {\bf q}, {\bf p}\big)}=q_x\quad\mbox{and}\quad  
  {\dot q_x\big(s;\nu, {\bf q}, {\bf p}\big)}=p_x,\quad x\in\bbZ_N
  \end{equation}
  for any $\nu\in\bbR$. Let  
  $
{\bf p}\big(t;\nu, {\bf q}, {\bf p}\big)=\Big(p_x\big(t;\nu, {\bf q}, {\bf p}\big)\Big),
$ where $p_x\big(t;\nu, {\bf q}, {\bf p}\big)=\dot q_x\big(t;\nu, {\bf q}, {\bf p}\big)$.
We omit writing the initial data $({\bf q}, {\bf p})$, if they are
obvious from the context.

We shall also denote by  
$$
{\bf q}_{\rm p}(t;\nu)=\Big(q_{x,{\rm p}}
(t;\nu)\Big)_{x\in\bbZ_N},\qquad {\bf p}_{\rm p}(t;\nu)=\Big(p_{x,{\rm p}}
(t;\nu)\Big)_{x\in\bbZ_N}
$$
a $\theta$-periodic solution  of \eqref{012805-24x}.

\subsection{Global stability of the periodic solution in the case of a harmonic
  chain on $\bbZ_N$}

\label{sec4.1}

In the case $\nu=0$ equation \eqref{012805-24y}, with the initial data at $s$ given by
$({\bf q},{\bf p})$, can be explicitly solved. Let us introduce some
auxiliary notation. Define a $2\times 2$ bloc matrix, with each block a $(2N+1)\times
  (2N+1)$ matrix,
  \begin{equation}
\label{AA}
A=
\left(
  \begin{array}{cc}
   {0_{2N+1}}&{\rm Id}_{2N+1}\\
    \Delta -\om_0^2 {\rm Id}_{2N+1}& -{\ga E_{2N+1}}
  \end{array}
\right),
\end{equation}
where ${\rm Id}_{n}$ is  the $n\times n$ identity matrix
and
$$
E_{2N+1}={[\delta_{-N-1}(x)\delta_{-N-1}(y)+\delta_{N+1}(x)\delta_{N+1}(y)]_{x,y=-N-1,\ldots,N+1}}.
$$
We denote by $\si(A)$ the set of all eigenvalues of $A$ and by ${\rm
  Re}\,\sigma(A)$ the set of all their real parts.
   Both here and in what follows by the norm of the matrix $M=[m_{i,j}]$ we
understand its operator norm
$
\|M\|:=\sup_{\|x\|=1}\|Mx\|,
$
where $\|\cdot\|$ is the euclidean norm of a vector.

Suppose that $({\bf q}, {\bf p})\in\Om_N$ and $s\le t$.
 Denote by ${\bf q}  (t,s )=\big(q_x(t,s)\big)$ the solution of \eqref{012805-24y}
satisfying
\begin{equation}
  \label{031602-25}
q_x  (s,s )=q_x,\quad \dot
  q_x  (s,s )=p_x,\quad x\in\bbZ_N.
\end{equation}
Using this 
notation   we can write
\begin{equation}
    \label{031111-24}
    \begin{split}
 &     \left(\begin{array}{c}
  \dot{\bf q} (t;s )\\
  \dot {\bf p}(t;s )
\end{array}
\right)=A  \left(\begin{array}{c}
   {\bf q}(t;s )\\
    {\bf p}(t;s )
\end{array}
\right) +  \mathbb
F\Big(\frac{t}{\theta}\Big) ,\\
&
{\bf q}(s;s)={\bf q},\quad {\bf p}(s;s)={\bf p}.
    \end{split}
  \end{equation}
The solution can be written as 
\begin{equation}
    \label{041111-24}
    \begin{split}
 &     \left(\begin{array}{c}
   {\bf q} (t;s )\\
    {\bf p}(t;s )
\end{array}
\right)=e^{A(t-s)}    \left(
  \begin{array}{c}
    {\bf q} \\
   {\bf p} 
\end{array}
\right) +\int_{s}^te^{A(t-s')}  \mathbb F\Big(\frac{s'}{\theta}\Big)\dd s',
    \end{split}
  \end{equation}
 { where
  $\mathbb F(t)$ is a vector valued function, with $2(2N+1)$ components that are all $0$,
  except the one corresponding to the momentum co-ordinate at $x=0$,
  where it equals ${\cal F}(t)$.}
  % \[
%   \begin{split}
%   \mathbb F(t) =
% \left(
%   \begin{array}{c}
%     0_{2N+1} \\
%     0_N\\
%    {{\cal F}(t)}\\
%     0_N
% \end{array}
% \right) .
% \end{split}
% \]
In particular,  $ {\bf q}_{\rm p}(t )$  the $\theta$-periodic solution of
  \eqref{012805-24x} is given by 
  \begin{equation}
    \label{051111-24}
    \begin{split}
      \left(
  \begin{array}{c}
    {\bf q}_{\rm p}(t )\\
   {\bf p}_{\rm p}(t ) 
\end{array}
\right)= \int_{-\infty}^te^{A(t-s)}  \mathbb
F\Big(\frac{s}{\theta}\Big)\dd s.
    \end{split}
  \end{equation}
 Define
\begin{equation}
  \label{la-N}
  \la_N:=-\lim_{t\to+\infty}\frac{\log\|e^{A t}\|_N}{t}.
  \end{equation}
It can be shown, see \cite[Proposition 1.2]{menegaki} that    there exists
$C_0>0$ such that
\begin{equation}
  \label{071111-24}
  \la_N \ge \frac{C_0}{N^{3}},\quad N=1,2,\ldots.
\end{equation}
As an immediate consequence of \eqref{la-N} we conclude the following.
 \begin{proposition}
    \label{cor011010-24}
    There exist $C_A>0$ such that
    \begin{equation}
      \label{121010-24}
      \|e^{At}\|\le C_Ae^{-\la_N t},\quad t\ge0.
    \end{equation}
    \end{proposition}
%From this we conclude the stability result.

   \medskip

 {We claim that, with the presence of dissipation on both ends of a
chain the solution $\big({\bf q}  (t,s,{\bf q},{\bf p}  ),\dot {\bf q}  (t,s,{\bf q},{\bf p}  )\big)$ tends to the
periodic solution as $s\to-\infty$.}
\begin{theorem}
   \label{thm011111-24}
  Suppose that $({\bf q}_{\rm p} (t  ))$ is the unique
  periodic solution given by \eqref{012110-24}. Then, {for $\gamma>0$}
  \begin{equation}
    \label{111010-24}
    \lim_{s\to-\infty}   \Big( \|{\bf q}  (t,s,{\bf q},{\bf p}  )-{\bf q}_{\rm p} (t
    )\|_N +\|\dot{\bf q}  (t,s,{\bf q},{\bf p}  )-\dot{\bf q}_{\rm p} (t
    )\|_N\Big)=0
  \end{equation}
  for any  $t\in\bbR$ and $({\bf q},{\bf p}  )\in\Om_N$.
   \end{theorem}
  \proof
   Using \eqref{041111-24} and \eqref{051111-24} we get
   \begin{equation}
    \label{061111-24}
       \left(\begin{array}{c}
   {\bf q} (t;s ) -{\bf q}_{\rm p}(t )\\
    {\bf p}(t;s ) -{\bf p}_{\rm p}(t )
\end{array}
\right)=e^{A(t-s)}  \Bigg[  \left(
  \begin{array}{c}
    {\bf q} \\
   {\bf p} 
\end{array}
\right)  -   \left(\begin{array}{c}
    {\bf q}_{\rm p}(s)\\
    {\bf p}_{\rm p}(s )
\end{array}
\right)\Bigg].
\end{equation}
The result is then an immediate conclusion from Proposition \ref{cor011010-24}.
   \qed

\subsection{Global stability via the approximation scheme}

  \label{sec5.3}
 \begin{theorem}
   \label{thm021010-24}
  Suppose that ${\bf q}_{\rm p} (t;\nu )$ is the
  periodic solution defined in Theorem \ref{main:thm} {and $\gamma>0$}.
    Then, for each
    $|\nu|< \nu_0$ (see \eqref{delta-m})  {and $\xi>0$ large enough}
  we have
 \begin{equation}
    \label{010201-25}
    \lim_{s\to-\infty}\int_0^{+\infty}e^{-\xi t}\|{\bf q}\big(t;s,\nu
    \big)  -{\bf q}_{\rm p}(t;\nu)\|^2\dd t=0.
    \end{equation}
  \end{theorem}
The proof of the theorem is given in Section \ref{sec6.4.2} below but
first we apply it to prove Theorem \ref{thm011111-24b}.

  \subsubsection*{Proof of Theorem \ref{thm011111-24b} in the case $|\nu|<\nu_0$}

  As a corollary of Theorem \ref{thm021010-24} we conclude that
$\Big({\bf q}\big(\cdot;s,\nu \big), {\bf p}\big(\cdot;s,\nu
\big)\Big)$ converges to the periodic solution $\big({\bf q}_{\rm p}(\cdot;\nu), {\bf p}_{\rm
  p}(\cdot;\nu)\big)$, as $s\to-\infty$, in $L^2_{\rm
  loc}[0,+\infty)$. Since  the functions are bounded, together with
their derivatives, the family is compact in the topology of uniform
convergence on compact subsets of $[0,+\infty)$. The $L^2_{\rm
  loc}$ convergence allows us to identify the limit, as the periodic
solution, which in particular implies the conclusion of  Theorem \ref{thm011111-24b}.\qed

\subsubsection{Approximation scheme}

 For fixed  $({\bf q},{\bf p})\in\Om_N$  and {$s <0$} 
define
$${\bf q}^{(L)}(t;s,{\bf q},{\bf p},\nu)=\Big(  q^{(L)}_{x}(t;s,{\bf q},{\bf p},\nu) \Big)_{x\in
  \bbZ_N}, \quad {t\ge 0},\quad L={0,}1,2,\ldots
$$ 
as the solution of
\begin{equation} 
\label{012805-24y0}
\begin{split}   
 & \ddot    q_{x}^{(0)}(t;s,\nu) =  \big(\Delta -\om_0^2\big) q_x^{(0)}(t;s,\nu) 
   -\ga\big(\delta_{-N}(x)+\delta_{N}(x)\big) \dot q_x^{(0)}(t;s,\nu)
   + {\cal F}(t/\theta)\delta_{0}(x) ,\\
    &
       q_{x}^{(0)}(0;s,\nu) =q_x(0;s, {\bf q},{\bf p},\nu)-\sum_{L=1}^{+\infty}q_{x,{\rm
           p}}^{(L)}(0;\nu)\nu^{L},\\
       &
       \dot    q_{x}^{(0)}(0;\nu) =\dot
       q_x(0;s, {\bf q},{\bf p},\nu)-\sum_{L=1}^{+\infty}\dot q_{x,{\rm p}}^{(L)}(0;\nu)\nu^{L},
       \quad
   x\in \bbZ_N
   \end{split}
 \end{equation}
and for $L=1,2,\ldots$ 
\begin{equation} 
\label{eq:flip-lZ}
\begin{aligned}
    \ddot    q_x^{(L)}(t;s,\nu) &=  \big(\Delta
   -\om_0^2\big) q_x^{(L)}(t;s,\nu) 
   -\ga\big(\delta_{-N,x}+\delta_{N,x}\big) \dot
   q_x^{(L)}(t;s,\nu)-   v_{x,L-1}(t;s,\nu) \\
    {q_{x}^{(L)}(0;s,\nu)} &= {q_{x,{\rm p}}^{(L)}(0;\nu)},\qquad
              \dot    q_{x}^{(L)}(0;s,\nu) = \dot q_{x,{\rm p}}^{(L)}(0;\nu)  ,\quad
  x\in \bbZ_N.
\end{aligned}
\end{equation}

Here
\begin{equation}
  \label{vxLZ}
  \begin{split}
&v_{x,L-1}(t;s,\nu)=\frac{1}{\nu^{L-1}}\Big[W_x\big({\bf Q}^{(L-1)}(t;s,\nu)\big)-W_x\big({\bf Q}^{(L-2)}(t;s,\nu)\big)\Big]    
\end{split}
\end{equation}
{where $W_x$ is defined by \eqref{W}} and
${\bf Q}^{(L)}(t;s,\nu)= \Big( Q_{x}^{(L)}(t;s,\nu)\Big)_{x\in \bbZ_N}$
is given by an analogue of \eqref{010905-24}:
\begin{equation}
  \label{010905-24-np}
  Q_{x}^{(L)}(t;s,\nu)=\sum_{\ell=0}^{L}q^{(\ell)}_{x}(t;s,\nu)\nu^\ell, \qquad t\ge 0, \qquad L\ge 0.
\end{equation} % with
% $q_{x,{\rm p}}^{(\ell)}(t;\nu)$ is replaced by  $q_x^{(\ell)}(t;s,\nu)$.
By convention $W_x\big({\bf Q}^{(-1)}(t;s,\nu)\big)\equiv 0$.
Both here and below we suppress writing the initial data ${\bf q},{\bf
  p}$ in the notation, when they are clear from the context.

{Notice that $\sum_{L=0}^{+\infty} q_{x}^{(L)}(0;s,\nu) \nu^L = q_x(0;s, {\bf q},{\bf p},\nu)$.}
Then for $L\ge 0$
\begin{equation} 
\label{012805-24L}
\begin{split}   &\ddot    Q_{x }^{(L)} (t;s,\nu)=  \big(\Delta
  -\om_0^2\big) Q_{x }^{(L)} (t;s,\nu) -\ga(
   \delta_{-N}(x)+
  \delta_N(x))\dot Q_{x }^{(L)}(t;s,\nu) 
   \\
   &
  \qquad    \qquad    \qquad    \qquad    \qquad  -\nu
   W\big(Q_{x }^{(L-1)}(t;s,\nu)\big)  
   + {\cal F}(t/\theta)\delta_{x,0} ,\\
   &
       Q_{x }^{(L)} (0;s,\nu) =q_x(0;s,\nu)-\sum_{\ell=L+1}^{+\infty}q_{x,{\rm
           p}}^{(\ell)}(0;\nu)\nu^{\ell}\\
   &
   \dot    Q_{x }^{(L)} (0;\nu) =q_x(0;s,\nu)-\sum_{\ell=L+1}^{+\infty}\dot q_{x,{\rm
           p}}^{(\ell)}(0;\nu)\nu^{\ell},\quad
   x\in \bbZ_N.
   \end{split}
 \end{equation}

 {Let
 $$
 {\frak Q}_L(t,s):=\|{\bf Q}^{(L)}(t,s; \nu)-{\bf
  Q}^{(L-1)}(t,s; \nu)\|+\|\dot{\bf Q}^{(L)}(t,s; \nu)-\dot{\bf
  Q}^{(L-1)}(t,s; \nu)\|
$$
for $L=1,2,\ldots$.
\begin{lemma}
   \label{lm010301-25}
   For fixed {$t_*>0> s$ and $|\nu|<\nu_0$} we have
   \begin{equation} 
     \label{010301-25}
        \sum_{L=1}^{+\infty}\sup_{t\in[0,t_*]}{\frak Q}_L(t,s) 
<+\infty.
     \end{equation}
   \end{lemma}
   \proof 
 The solutions of \eqref{012805-24L}  satisfy
\begin{equation}
    \label{121111-24z}
    \begin{split}
 &     \left(\begin{array}{c}
 {\bf  Q} ^{(L)} (t;s,\nu)- {\bf  Q} ^{(L-1)} (t;s,\nu)\\
   \dot {\bf  Q} ^{(L)} (t;s,\nu)-\dot {\bf  Q} ^{(L-1)} (t;s,\nu)
\end{array}
\right)=\nu^Le^{A t}    \left(
  \begin{array}{c}
   {\bf q}_{{\rm
           p}}^{(L)}(0;\nu)\\
  \dot{\bf q}_{{\rm
           p}}^{(L)}(0;\nu)
\end{array}
\right)\\
&
-\nu\int_{0}^te^{A(t-s')}  \Big[ \mathbb V\big( {\bf  Q} ^{(L-1)} (s';s,\nu)\big)-\mathbb V\big({\bf  Q} ^{(L-2)} (s';s,\nu)\Big)\Big]
\dd s'
    \end{split}
  \end{equation}
  for any $L=1,2,\ldots$, $t\ge0$.  
 Here $\mathbb V:\Om_N\to \Om_N\times \Om_N$ and $ {\cal  W}:\Om_N\to
   \Om_N$ are
functions defined as
\begin{equation}
  \label{bV}
  \mathbb V(f)  =
\left(
  \begin{array}{c}
    0 \\
  {\cal  W}(f) ,
\end{array}
\right),\qquad 
{\cal W}_x(f)=W_x(f),\quad x\in\bbZ_N
\end{equation}
(cf \eqref{W})   and, by convention,
$
\mathbb V\big( {\bf  Q} ^{(-1)} (s';s,\nu)\big)\equiv 0.
$
Hence
\begin{align}
  \label{011003-25}
  &
    {\frak Q}_L(t,s)\le  |\nu|^L\big(\|{\bf q}_{{\rm
           p}}^{(L)}(\cdot;\nu)\|_{\infty}+\|\dot {\bf q}_{{\rm
           p}}^{(L)}(\cdot;\nu)\|_{\infty}\big)+ |\nu|{\frak
    V}\int_{0}^t {\frak Q}_{L-1}(s',s)\dd s'\notag\\
  &
  \le   \ldots\le |\nu|^L\big(\|{\bf q}_{{\rm
           p}}^{(L)}(\cdot;\nu)\|_{\infty}+\|\dot {\bf q}_{{\rm
           p}}^{(L)}(\cdot;\nu)\|_{\infty}\big)+|\nu|^{L-1}\big(\|{\bf q}_{{\rm
           p}}^{(L-1)}(\cdot;\nu)\|_{\infty}+\|\dot {\bf q}_{{\rm
           p}}^{(L-1)}(\cdot;\nu)\|_{\infty}\big)\frac{|\nu|{\frak
    V}t}{1!}\\
  &
    +\ldots +  \big(\|{\bf q}_{{\rm
           p}}^{(0)}(\cdot;\nu)\|_{\infty}+\|\dot {\bf q}_{{\rm
           p}}^{(0)}(\cdot;\nu)\|_{\infty}\big)\frac{(|\nu|{\frak
    V}t)^L}{L!} 
    .\notag
\end{align}
Summing over $L$ we conclude that
\begin{align}
  \label{011003-25}
  &
   \sum_{L=1}^{+\infty} \sup_{t\in[0,t_*]}{\frak Q}_L(t,s)\le \Big(\sum_{L=0}^{+\infty} |\nu|^L\big(\|{\bf q}_{{\rm
           p}}^{(L)}(\cdot;\nu)\|_{\infty}+\|\dot {\bf q}_{{\rm
           p}}^{(L)}(\cdot;\nu)\|_{\infty}\big)\Big)\exp\left\{|\nu|{\frak
    V}t_*\right\} 
    .\notag
\end{align}}
The conclusion of the lemma then follows from Proposition \ref{prop012410-24}.
   \qed

\medskip
   
As a consequence of Lemma \ref{lm010301-25} we have {for $t>0$}
 \begin{equation} 
\label{011112-24}
\begin{split}
 &\sum_{\ell=0}^{+\infty}q^{(\ell)}_{x}(t;s,\nu)\nu^\ell= \lim_{L
   \to+\infty}Q_{x}^{(L)}(t;s,\nu) =q_{x}(t;s,\nu),\\
 &
 \sum_{\ell=0}^{+\infty}\dot q^{(\ell)}_{x}(t;s,\nu)\nu^\ell= \lim_{L
   \to+\infty}\dot Q_{x}^{(L)}(t;s,\nu) =\dot q_{x}(t;s,\nu).
   \end{split}
 \end{equation}

\subsubsection{Approximation scheme for the Laplace transform}
 Consider the Laplace transforms
 \[
      \hat q_x^{(L)}(\la;s,\nu)=\int_0^{+\infty}e^{-\la t}
      q_x^{(L)}(t;s,\nu) \dd t.
    \]
    They satisfy the system
     \begin{equation} 
   \label{eq:HZ-0-L0}
   \begin{split}
  &0 =  \left[\Delta  
  -\om_0^2 -\la^2-\ga \la(
    \delta_{-N}(x)+
   \delta_N(x))\right]  \hat q_{x}^{(0)}(\la;s,\nu ) \\
  &
  +\big[\la+\ga  (
    \delta_{-N}(x)+
    \delta_N(x))\big]    q_{x}^{(0)}(0;s,\nu ) + \dot
   q_{}^{(0)}(0;s,\nu ) 
  +  \;    \delta_{x,0} \sum_{m\in\bbZ}\frac{ \hat{\rm  F}_m }{{\la-im\om }}  
    , \; \quad x\in\bbZ_N
  \end{split}
\end{equation}
      and for $L=1,2,\ldots$  
\begin{equation} 
\label{011005-21-0l-L}
\begin{aligned}
  0 &=  \left[\Delta  
  -\om_0^2 -\la^2-\ga \la(
    \delta_{-N}(x)+
   \delta_N(x))\right]  \hat q_{x}^{(L)}(\la;s,\nu ) \\
  &
  +\big[\la+\ga  (
    \delta_{-N}(x)+
    \delta_N(x))\big]    q_{x,{\rm p}}^{(L)}(0;\nu ) + \dot
   q_{x,{\rm p}}^{(L)}(0;\nu ) 
   {-}  \;     \hat v_{x,L-1}(\la;s,\nu)
    , \; \quad x\in\bbZ_N,
  \end{aligned} \end{equation}
with
\begin{equation}
  \label{022810-24-L}
 \hat v_{x,L-1} (\la;s,\nu)=  \int_0^{+\infty} e^{-\la t}
v_{x,L-1}  (t;s,\nu) \dd t,
\end{equation}
 {where $v_{x,L-1}(t;s,\nu)$ is given by   \eqref{vxLZ}.}

Let $\hat q_{x,{\rm p}}^{(L)}(\la;\nu )$ be the Laplace transform of
$q_{x,{\rm p}}^{(L)}(t;\nu )$ and 
\begin{align}
  \label{060201-25}
  &
    \delta  q_{x}^{(L)}(t;s ,\nu ):=    q_{x}^{(L)}(t;s,\nu
    )- q_{x,{\rm p}}^{(L)}(t; \nu ),\notag\\
  &
 \delta\hat q_{x}^{(L)}(\la;s ,\nu ):=  \hat q_{x}^{(L)}(\la;s,\nu
    )-\hat q_{x,{\rm p}}^{(L)}(\la;\nu ),\\
    &
    \delta  Q_{x}^{(L)}(t;s ,\nu ):=    Q_{x}^{(L)}(t;\nu
    )- Q_{x,{\rm p}}^{(L)}(t;\nu ). \notag
\end{align}
For $L=1,2,\ldots$ we get
 \begin{equation} 
\label{021005-21-0l-L}
\begin{aligned}
 & 0 =  \left[\Delta  
  -\om_0^2 -\la^2-\ga \la(
    \delta_{-N}(x)+
   \delta_N(x))\right] \delta \hat q_{x}^{(L)}(\la;s,\nu )  
    {-}  \;     \delta\hat v_{x,L-1}(\la;s,\nu)
    , \; \quad x\in\bbZ_N,\quad\mbox{with}
    \\
    &
    \delta\hat v_{x,L-1}(\la;s,\nu):= \hat v_{x,L-1}(\la;s,\nu)-\hat
    v_{x,L-1}^{({\rm p})}(\la;\nu) .
  \end{aligned} \end{equation}
Let $\la=\xi+i\eta$, where $\xi>0$ and $\eta\in\bbR$.
Multiplying both sides by $\Big(\delta \hat q_{x}^{(L)}(\la;s,\nu )
\Big)^\star$ and summing over $x$ we get
\begin{equation} 
  \label{041112-24}
  \begin{split}
   &\left(\om_0^2+\xi^2-\eta^2 +2i\xi\eta\right) 
 \sum_{x\in\bbZ_N}\big|\delta \hat q_{x}^{(L)}(\la;s,\nu )\big|^2
 +\sum_{x\in\bbZ_N}\big|\nabla_x\delta \hat q_{x}^{(L)}(\la;s,\nu
 )\big|^2\\
 &
 +\ga \left(\xi+i\eta\right) \Big[\big|\delta \hat
 q_{-N}^{(L)}(\la;s,\nu ) \big|^2+\big|\delta \hat q_{N}^{(L)}(\la;s,\nu )
 \big|^2\Big]\\
 &=
  {-}\sum_{x\in\bbZ_N} \delta\hat v_{x,L-1}(\la; s,\nu) \Big(\delta \hat q_{x}^{(L)}(\la;s,\nu )
   \Big)^\star  .
   \end{split}
    \end{equation}
  Comparing the real and imaginary parts we conclude that
    \begin{equation} 
  \label{041112-24a}
  \begin{split}
   &\left(\om_0^2+\xi^2-\eta^2  \right) 
 \sum_{x\in\bbZ_N}\big|\delta \hat q_{x}^{(L)}(\la;s,\nu )\big|^2
 +\sum_{x\in\bbZ_N}\big|\nabla_x\delta \hat q_{x}^{(L)}(\la;s,\nu
 )\big|^2 
 \\
 &
 +\ga  \xi \Big[\big|\delta \hat
 q_{-N}^{(L)}(\la;s,\nu ) \big|^2+\big|\delta \hat q_{N}^{(L)}(\la;s,\nu )
 \big|^2\Big]= {-}\sum_{x\in\bbZ_N} {\rm Re}\,\Bigg[\delta\hat v_{x,L-1}(\la;s,\nu) \Big(\delta \hat q_{x}^{(L)}(\la; s,\nu )
 \Big)^\star  \Bigg]
 \end{split}
    \end{equation}
    and
    \begin{equation} 
  \label{041112-24b}
  \begin{split}
 &
 2\xi\eta
 \sum_{x\in\bbZ_N}\big|\delta \hat q_{x}^{(L)}(\la;s,\nu )\big|^2
 +\ga  \eta \Big[\big|\delta \hat
 q_{-N}^{(L)}(\la;s,\nu ) \big|^2+\big|\delta \hat q_{N}^{(L)}(\la;s,\nu )
 \big|^2\Big]\\
 &
 ={-} \sum_{x\in\bbZ_N} {\rm Im}\,\Bigg[\delta\hat v_{x,L-1}(\la; s,\nu) \Big(\delta \hat q_{x}^{(L)}(\la;s,\nu )
   \Big)^\star  \Bigg].
      \end{split}
    \end{equation}
    {Recall that $\delta_* < \omega_0^2$. Choose $a\in (\delta_*,\omega_0^2)$.}
    For $\eta^2\le \om_0^2-a$ and all $\xi>0$
    we conclude from \eqref{041112-24a} that
   \begin{equation} 
  \label{030201-25}
  \begin{split}
a\sum_{x\in\bbZ_N}\big|\delta \hat q_{x}^{(L)}(\la;s,\nu )\big|^2 \le 
   \Big\{\sum_{x\in\bbZ_N}\Big[\delta\hat
 v_{x,L-1}(\la; s,\nu) 
    \Big]^2\Big\}^{1/2}\Big\{\sum_{x\in\bbZ_N}\big|\delta \hat q_{x}^{(L)}(\la; s,\nu )\big|^2\Big\}^{1/2}.
 \end{split}
    \end{equation}
    For $\eta\ge \sqrt{\om_0^2-a}$  we get from \eqref{041112-24b} that 
 \begin{equation} 
  \label{041112-24c}
  \begin{split}
2\xi \sqrt{\om_0^2-a} \sum_{x\in\bbZ_N}\big|\delta \hat q_{x}^{(L)}(\la;s,\nu )\big|^2
 \le  \Big\{\sum_{x\in\bbZ_N}\Big[\delta\hat v_{x,L-1}(\la; s,\nu) \Big]^2\Big\}^{1/2}
 \Big\{\sum_{x\in\bbZ_N}\big|\delta \hat q_{x}^{(L)}(\la; s,\nu )\big|^2\Big\}^{1/2}.
      \end{split}
    \end{equation}
    If $\eta\le -\sqrt{\om_0^2-a}$, then we multiply both sides of
      \eqref{041112-24b}   by $-1$
   and \eqref{041112-24c} is still in force.
      Consequently choosing $\xi$ such that  $2\xi \sqrt{\om_0^2-a}\ge a$
      we have that \eqref{030201-25} holds for any $\eta$.

From the above consideration  we conclude the following.
\begin{lemma}
  \label{lm010201-25}
  {For $\la=\xi+i\eta$, $\eta\in\bbR$, 
  $\delta_*< a < \omega_0^2$ and $\xi>\frac{a}{2\sqrt{\omega_0^2- a}}$
  we have}
    \begin{equation} 
  \label{040201-25}
  \begin{split}
{a^2}  \sum_{x\in\bbZ_N}\big|\delta \hat q_{x}^{(L)}(\la;s,\nu
)\big|^2  \le 
    \sum_{x\in\bbZ_N}\Big[\delta\hat
 v_{x,L-1}(\la; s,\nu) 
    \Big]^2 .
 \end{split}
\end{equation}
  \end{lemma}

  \medskip

  \begin{corollary}
    \label{cor010201-25}
    For $a$ and $\xi$ as in the statement of Lemma \ref{lm010201-25}  we have
    \begin{equation} 
    \begin{split}
      \label{070201-25}
&{a^2}
\sum_{x\in\bbZ_N}\int_0^{+\infty}e^{-2\xi t}\big[\delta  
q_{x}^{(L)}(t;s,\nu )\big]^2 \dd t\\
&
\le 
  {\frak V}^2   \sum_{x\in\bbZ_N}\int_0^{+\infty}e^{-2\xi t}[\delta
  q^{(L-1)}_{x}(t;s,\nu)]^2\dd t ,\quad L=1,2,\ldots.
 \end{split}
\end{equation}
    \end{corollary}
    \proof
    Integrating over $\eta$ in \eqref{040201-25}
    and using the Parseval identity
    we conclude that
\begin{equation}
  \label{080201-25}
   \begin{split}
{a^2}
\sum_{x\in\bbZ_N}\int_0^{+\infty}e^{-2\xi t}\big[\delta  
q_{x}^{(L)}(t;s,\nu )\big]^2 \dd t 
\le 
    \sum_{x\in\bbZ_N}\int_0^{+\infty}e^{-2\xi t}\Big[\delta 
 v_{x,L-1}(t; s,\nu) 
    \Big]^2 \dd t .
 \end{split}
\end{equation}
Using \eqref{vxLZ} and \eqref{Vp1z} we conclude in the same way as in \eqref{110301-25} that
\begin{align*}
& |\delta   v_{x,L-1}(t; s,\nu)|
    \le (\|V''\|_{\infty}+\|U''\|_{\infty})  |\delta q^{(L-1)}_{x}(t;s,\nu)|\\
  &
    +\|U''\|_{\infty}\Big(| \delta q^{(L-1)}_{x-1}(t;s,\nu)|+|\delta  q^{(L-1)}_{x+1}(t;s,\nu)|\Big).
\end{align*}
Hence,
\begin{align}
  \label{102410-24s}
& [ \delta v_{x,L-1}(t; s,\nu)]^2 
    \le (\|V''\|_\infty+\|U''\|_{\infty})^2
                      [\delta q^{(L-1)}_{x }(t;s,\nu)]^2\notag\\
  &
    +(\|V''\|_\infty+\|U''\|_{\infty}) \|U''\|_{\infty}\Big( 2[\delta q^{(L-1)}_{x}(t;s,\nu)]^2+ [\delta q^{(L-1)}_{x-1}(t;s,\nu)]^2+ [\delta q^{(L-1)}_{x+1}(t;s,\nu)]^2\Big)\\
  &
    +2 \|U''\|_{\infty}^2\Big([\delta  q^{(L-1)}_{x-1}(t;s,\nu)]^2+[
    \delta q^{(L-1)}_{x+1}(t;s,\nu)]^2\Big) 
    .\notag
\end{align}
Combining with \eqref{080201-25}
  we conclude estimate \eqref{070201-25} with $\frak{V}$ defined in \eqref{delta-m}.
 
    \qed

    \subsubsection{Proof of Theorem \ref{thm021010-24}}
    \label{sec6.4.2}
   % We start with the following.
  %   \begin{lemma}
%       \label{lm020301-25}
%       For fixed $s$, $({\bf q},{\bf p})\in\bbR^{2(2N+1)}$ and $\nu$ we have
%       \begin{equation}
%     \label{040301-25}
%   \sup_{t\ge s}   \Big( \|{\bf q}  (t,s ;\nu)\| +\|{\bf p}  (t,s ;\nu )\|\Big)=Q(s ,{\bf q},{\bf p})<+\infty.
%   \end{equation}
%       \end{lemma}
%       \proof
%       Using \eqref{121111-24} we conclude that
% \begin{equation}
%     \label{050301-25}
%     \begin{split}
%  &     \Big( \|{\bf q}  (t,s ;\nu)\|^2 +\|{\bf p}  (t,s ;\nu )\|^2\Big)^{1/2}\le
%  \|e^{A(t-s)}\| \Big( \|{\bf q} \|^2 +\|{\bf p}  \|^2\Big)^{1/2}\\
%  &
%  +|\nu|\int_{s}^t\|e^{A(t-s')}\|\|   \mathbb V( {\bf q} (s',s;\nu ))\|_{\ell_2(\bbZ_N)}
% \dd s'\\
% &
% +\int_{s}^t\|e^{A(t-s')}\| \|\mathbb
% F\Big(\frac{s'}{\theta}\Big) \|_{\ell_2(\bbZ_N)}\dd s'.
%     \end{split}
%   \end{equation}
%  Thanks to  \eqref{071111-24} we can write
%   \begin{equation}
%     \label{060301-25}
%     \begin{split}
%  &     \Big( \|{\bf q}  (t,s ;\nu)\|^2 +\|{\bf p}  (t,s ;\nu )\|^2\Big)^{1/2}\le
%  C e^{-\la_N(t-s)} \Big( \|{\bf q} \|^2 +\|{\bf p}  \|^2\Big)^{1/2}\\
%  &
%  +C \int_{s}^t e^{-\la_N(t-s')} 
% \dd s' .
%     \end{split}
%   \end{equation}
%   Hence \eqref{040301-25} follows.
%       \qed

%       \medskip

 We start with the following.
    \begin{lemma}
      \label{lm020301-25}
      For fixed  $({\bf q},{\bf p})\in\Omega_N$ and $\nu$ we have
      \begin{equation}
    \label{040301-25}
    \sup_{t\ge s}   \Big( \|{\bf q}  (t,s ; {\bf q},{\bf p},\nu)\|
    +\|{\bf p}  (t,s ; {\bf q},{\bf p},\nu)\|\Big)
    :=  {\frak Q}({\bf q},{\bf p})<+\infty.
  \end{equation}
      \end{lemma}
      \proof
     The solution of \eqref{012805-24x}  satisfies
\begin{equation}
    \label{121111-24}
    \begin{split}
 &     \left(\begin{array}{c}
   {\bf q} (t,s;\nu )\\
    {\bf p}(t,s;\nu )
\end{array}
\right)=e^{A(t-s)}    \left(
  \begin{array}{c}
    {\bf q} \\
   {\bf p} 
\end{array}
\right)-\nu\int_{s}^te^{A(t-s')}   \mathbb V( {\bf q} (s',s;\nu ))
\dd s'\\
&
+\int_{s}^te^{A(t-s')}  \mathbb
F\Big(\frac{s'}{\theta}\Big)\dd s'.
    \end{split}
  \end{equation}
 Here $\mathbb V:\Om_N\to \Om_N\times \Om_N$ is a
function defined in \eqref{bV}.
% \begin{equation}
%   \label{bV}
%   \mathbb V(f)  =
% \left(
%   \begin{array}{c}
%     0 \\
%   {\cal  W}(f) ,
% \end{array}
% \right),\qquad 
% {\cal W}_x(f)=W_x(f),\quad x\in\bbZ_N,
% \end{equation}
% cf \eqref{W}. 
      Using \eqref{121111-24} we conclude that
\begin{equation}
    \label{050301-25}
    \begin{split}
 &     \Big( \|{\bf q}  (t,s ;\nu)\|^2 +\|{\bf p}  (t,s ;\nu )\|^2\Big)^{1/2}\le
 \|e^{A(t-s)}\| \Big( \|{\bf q} \|^2 +\|{\bf p}  \|^2\Big)^{1/2}\\
 &
 +|\nu|\int_{s}^t\|e^{A(t-s')}\|\|   \mathbb V( {\bf q} (s',s;\nu ))\|_{\ell_2(\bbZ_N)}
\dd s'\\
&
+\int_{s}^t\|e^{A(t-s')}\| \|\mathbb
F\Big(\frac{s'}{\theta}\Big) \|_{\ell_2(\bbZ_N)}\dd s'.
    \end{split}
  \end{equation}
 Thanks to  \eqref{071111-24} we can write
  \begin{equation}
    \label{060301-25}
    \begin{split}
 &     \Big( \|{\bf q}  (t,s ;\nu)\|^2 +\|{\bf p}  (t,s ;\nu )\|^2\Big)^{1/2}\le
 C e^{-\la_N(t-s)} \Big( \|{\bf q} \|^2 +\|{\bf p}  \|^2\Big)^{1/2}\\
 &
 +C \int_{s}^t e^{-\la_N(t-s')} 
\dd s' .
    \end{split}
  \end{equation}
  Hence \eqref{040301-25} follows.
      \qed

      \medskip

  Consider the equation
\begin{equation} 
\label{012805-24y00}
\begin{split}   
 &    \ddot    {\frak q}_{x}(t) =  \big(\Delta
   -\om_0^2\big)  {\frak q}_{x}(t) 
   -\ga\big(\delta_{-N}(x)+\delta_{N}(x)\big)  \dot
   {\frak q}_{x}(t) ,\quad x=-N,\ldots,N.
   \end{split}
 \end{equation}
Denote by $B_R$ the ball of radius $R>0$, centered
      at $0$ in $\bbR^{2(2N+1)}$ and by ${\frak q}(t)=\big({\frak
        q}_x(t)\big)$, $\dot{\frak q}(t)=\big(\dot{\frak
        q}_x(t)\big)$.
      \begin{lemma}
      \label{lm030301-25}
      For fixed   $R>0$, $\nu$ and $\xi\ge 0$ we have
      \begin{equation}
    \label{060301-25}
  \lim_{s\to-\infty}\sup_{({\frak q}(0) , \dot{\frak q}(0))\in B_R}   \sum_{x\in\bbZ_N}\int_0^{+\infty}e^{-2\xi t}{\frak q}_x^2(t)\dd t=0.
  \end{equation}
\end{lemma}
\proof
 Using \eqref{121010-24} we conclude that for $t\ge 0\ge s$ we have
\begin{equation}
    \label{070301-25}
    \begin{split}
      \left(\sum_{x\in\bbZ_N} {\frak q}_x^2(t)\right)^{1/2} 
      \le Ce^{-\la_Nt}   {\frak Q}({\bf q},{\bf p}) .
\end{split}
\end{equation}
From this
\begin{equation}
    \label{080301-25}
     \sum_{x\in\bbZ_N} \int_0^{+\infty}e^{-2\xi t}{\frak q}_x^2(t) \dd t 
       \le  C {\frak Q}^2({\bf
         q},{\bf p}) \int_0^{+\infty}e^{-2\la_N t} \dd t   
       =\frac{C}{2\la_N} {\frak Q}^2({\bf q},{\bf p})   
      \end{equation}
  and \eqref{060301-25} follows.
\qed

 \medskip

      We restore the dependence on the initial data $({\bf q},{\bf
        p})\in\bbR^{2(2N+1)}$ in the notation of $\delta  
q_{x}^{(0)}(t;s, {\bf q},{\bf
  p},\nu )$.
From Lemmas \ref{lm020301-25} and \ref{lm030301-25} we immediately
conclude the following.
      \begin{corollary}
        \label{cor030301-25}
        We have
      \begin{equation}
        \label{060301-25}
  \lim_{s\to-\infty} \sum_{x\in\bbZ_N}\int_0^{+\infty}e^{-2\xi t}\big[\delta  
  q_{x}^{(0)}(t;s, {\bf q},{\bf   p},\nu )\big]^2 \dd t=0.
  \end{equation}
\end{corollary}
\proof
Note that $\delta  
q^{(0)}(0;s, {\bf q},{\bf
  p},\nu )=\big(\delta  
  q_{x}^{(0)}(t;s, {\bf q},{\bf   p},\nu )\big)_{x\in\bbZ_N}$ solves
  \eqref{012805-24y00}. According to Lemma \ref{lm020301-25} there
  exists $R>0$ such that
  $$
  \big(\delta  
q^{(0)}(0;s, {\bf q},{\bf
  p},\nu ), \delta  
\dot q^{(0)}(0;s, {\bf q},{\bf
  p},\nu )\big)\in B_R\quad\mbox{ for all }s<0.
$$
  The conclusion of the corollary then follows directly from  Lemma  \ref{lm030301-25}.\qed

\medskip
{Notice that $\frac{{\frak V} }{a} < \frac{{\frak V} }{\delta_*} :=\nu_0$.
Suppose that $|\nu|<\nu_0$.
% Choose $a\in \Big(\delta_*{|\nu|/\nu_0},\om_0^2)$.
Using Corollary  \ref{cor010201-25} we conclude that for $\xi$ large enough % given
% by formula \eqref{100301-25} 
we have}
\begin{equation}
  \begin{split}
      \label{070301-25}
      \left\{\sum_{x\in\bbZ_N}\int_0^{+\infty}e^{-2\xi t}
        \big[\delta q_{x}^{(L)}(t;s,\nu )\big]^2\right\}^{1/2}
\dd t
\le 
 \Big(\frac{{\frak V} }{a}\Big)^{{L}} \left\{\int_0^{+\infty}e^{-2\xi t}[\delta
  q^{(0)}_{x}(t;s,\nu)]^2\dd t\right\}^{1/2} ,\quad L=1,2,\ldots.
 \end{split}
\end{equation}
Hence,
\begin{equation}
  \begin{split}
    &\left\{ \sum_{x\in\bbZ_N}  \int_0^{+\infty}e^{-2\xi t}\big[q_{x}(t;s,\nu )
      -q_{x,{\rm p}}(t,\nu )\big]^2 \dd t \right\}^{1/2} \\
   &= \left\{\sum_{x\in\bbZ_N} \int_0^{+\infty}e^{-2\xi t} \big[
  \sum_{\ell\ge 0} \nu^\ell \delta q^{(\ell)}_{x}(t;s,\nu )\big]^2 \dd t \right\}^{1/2}  \\
   & \le  \sum_{\ell\ge 0}  |\nu|^\ell \left\{\int_0^{+\infty}e^{-2\xi t}\sum_{x\in\bbZ_N} \big[
   \delta q^{(\ell)}_{x}(t;s,\nu )\big]^2 \dd t \right\}^{1/2} \\
      \label{080301-25}
% \Bigg\{\sum_{x\in\bbZ_N}\int_0^{+\infty}e^{-2\xi_0 t}\big[ 
% q_{x}(t;s,\nu )-q_{x,{\rm p}}(t,\nu )\big]^2 \dd t\Bigg\}^{1/2}
&  \le 
 \frac{1}{1-(|\nu|{\frak V} /a)} \Bigg\{\int_0^{+\infty}e^{-2\xi t}[\delta
 q^{(0)}_{x}(t;s,\nu)]^2\dd t \Bigg\}^{1/2} .
 \end{split}
\end{equation}
% Here 
% $$
% K:=\Big[\Big({\bf q}(t;s ,\nu)-{\bf q}_{\rm
%   p}(t;\nu)+{\bf q}^{(0)}_{{\rm
%            p}}(0;\nu),
% \dot{\bf q}(t;s ,\nu)-\dot{\bf q}_{\rm
%   p}(t;\nu) +\dot {\bf q}^{(0)}_{{\rm
%            p}}(0;\nu)\Big),\quad t\ge s\Big].
% $$
The conclusion of Theorem \ref{thm021010-24} then follows from  Corollary   \ref{cor030301-25}. \qed

\appendix

\section{Formulas for the Green's function}

\subsection{Green's function of the operator $\la+\om_0^2-\Delta$ on $\bbZ$}

\label{secA1}

Suppose that $-\lambda \in \mathbb C\setminus{\cal I}$. The Green's function of  the operator $\la-\om_0^2+\Delta $, where
$\Delta$ is the free Laplacian on $\bbZ$ is given by
\begin{align}
  \label{Gla}
 {\rm G}_\la(x) = \left(\la+ \om^2_0 -\Delta\right)^{-1}{\delta_{0}}(x)\quad 
    =\int_0^1\frac{\cos(2\pi
                 ux)\dd u}{\la +4\sin^2(\pi u)+\om_0^2}.
\end{align}
Note that
\begin{align}
  \label{Gla1}
 {\rm G}_{\la^\star}(x) = {\rm G}_{\la}^\star(x) ,\quad x\in\bbZ,\, -\lambda \in \mathbb C\setminus{\cal I}.
\end{align}
Let  
$
\zeta=\chi(\la):=\frac12(2+\om_0^2+\la)
$ and let $\Phi_+$ be the inverse of the Joukowski function
$$
\zeta=J(z)=\frac12\Big(z+\frac1z\Big),\quad z\in\mathbb C
$$
considered for $|z|>1$. Then
$
z=\Phi_+(\zeta)$,  $\zeta\not\in[-1,1]$.
Condition
  $-\la\not\in {\cal I}$  is equivalent with
  $\zeta\not\in[-1,1]$.  Furthermore $|\Phi_+\big(\chi(\la)\big)|>1$.
 We have then
  \begin{equation}
    \label{012802-23}
  {\rm G}_\la(x)
  =\frac{\Phi_+^{1-|x|}\big(\chi(\la)\big)}{\Phi_+^2\big(\chi(\la)\big)-1}=\frac{\Phi_+^{-|x|}\big(\chi(\la)\big)}{2[\Phi_+\big(\chi(\la)\big)-\chi(\la)]},\quad
  x\in\bbZ
\end{equation}
We can use the mapping $\zeta\mapsto\Phi_+(\zeta)$,  $\zeta\not\in[-1,1]$ to define the
mapping $\zeta\mapsto w= (\zeta^2-1)^{1/2}$,   $\zeta\not\in[-1,1]$ by
letting
\begin{equation}
  \label{sqt}
(\zeta^2-1)^{1/2}:=\Phi_+(\zeta)-\zeta,\quad \zeta\not\in[-1,1].
\end{equation}
  In
particular,
$$
\{\zeta^2-1\}^{1/2}=\left\{
  \begin{array}{ll}
    \sqrt{\zeta^2-1},&\mbox{when }\zeta\in[1,+\infty),\\
    i\sqrt{1-\zeta^2},&\mbox{when }\zeta\in i(0,+\infty) ,\\
    -\sqrt{\zeta^2-1},&\mbox{when }\zeta\in(-\infty,-1],\\
    -  i\sqrt{1-\zeta^2},&\mbox{when }\zeta\in i(-\infty,0)
         \end{array}\right.
$$
Here $w\mapsto \sqrt{w}$ is  the branch of the inverse of $z\mapsto
z^2$ such that ${\rm Re}\, \sqrt{w}>0$, when $w\in\mathbb
C\setminus(-\infty,0]$.
   Furthermore
  \begin{align}
    \label{011801-24z}
   \Phi_+\big(\chi(\la)\big) =\zeta+\{\zeta^2-1\}^{1/2}=\Phi_+(\zeta).\notag
  \end{align}
Formula \eqref{012802-23}   can be rewritten in the form
\begin{equation}
    \label{012802-23zz}
  {\rm G}_\la(x)
  =\frac{\left\{1+\frac12(\la+\om_0^2)
      +\frac12\{(\la+\om_0^2)(\la+\om_u^2)\}^{1/2}\right\}^{-|x|}}{\{(\la+\om_0^2)(\la+\om_u^2)\}^{1/2}},\quad
  x\in\bbZ
\end{equation}
recall that $\om_u=\sqrt{\om_0^2+4}$.

Suppose that $\om\in\bbR$.
   In case $|m|\om<\om_0$ we have
  \begin{align}
    \label{021801-24n}
    &  G_{-(m\om)^2}(x)=\frac{ \xi_+^{-|x|} (m\om)}{D(m\om)},\quad
      x\in\bbZ,\quad\mbox{with}\notag\\
    &
      \xi_+(\Om):=1+\frac12\big( \om_0^2 -\Om^2+ D(\Om)\big)\quad\mbox{and}\\
    &
     D(\Om):= \sqrt{(\om_0^2 -\Om^2)(\om_u^2 -\Om^2)},\quad \Om\in [\om_0,\om_u]\notag
  \end{align}
  When $|m|\om>\om_u$ we have
\begin{align}
    \label{021801-24nn}
    &  G_{-(m\om)^2}(x)=-\frac{ \xi_-^{-|x|} (m\om)}{D(m\om)},\quad
      x\in\bbZ,\quad\mbox{with}\notag\\
    &
      \xi_-(\Om):=1+\frac12\big( \om_0^2 -\Om^2- D(\Om)\big),\quad
  x\in\bbZ. 
  \end{align}
  We have
  \begin{align*}
    &
      \xi_+(\Om)>1,\quad\mbox{when}\quad  0\le \Om<\om_0\quad\mbox{and}\\
    &
     \xi_-(\Om)<-1,\quad\mbox{when}\quad \Om>\sqrt{4+\om^2_0}.
  \end{align*}

\subsection{Green's function on $\bbZ_N$}
Suppose that
$G_\la^{(N)}(x,y)=\big(\la+\om_0^2-\Delta\big)^{-1}\delta_y(x)$,
$x,y\in\bbZ_N$ is the Green's function for  the Neumann laplacian on $\bbZ_N$. 
We have
 \begin{equation}
   \label{020107-23}
  G_\la^{(N)}(x,y)= \sum_{j=0}^{2N}
\frac{\psi_j(x) \psi_j(y) }{\la+\om_j^2},\quad x,y\in\bbZ_N.
   \end{equation}
Here $ \mu_j=\om_j^2$, $j=0,\ldots,2N$ are 
   the eigenvalues  of $\om_0^2-\Delta $, with
   \begin{equation}
     \label{omj}
        \omega_j = \omega\left(\frac{j}{2N+1}\right) 
         \end{equation}
         and the dispersion relation $\om(k)=\sqrt{\om_0^2+4\sin^2(\pi
           k)}$. The
   eigenvectors are given by
   \begin{equation}
     \label{psij}
     \begin{split}
&\psi_0(x)=\left(\frac{1}{2N+1}\right)^{1/2},\\
&
\psi_j(x)=\left(\frac{2}{2N+1}\right)^{1/2}\cos\Big(\frac{\pi
  j}{2}\cdot \frac{ 2(x+N)+1}{2N+1}\Big) \quad   x=-N,\ldots,N,\,j=1,\ldots,2N.
\end{split}
\end{equation}

Using the method of images we obtain the following formula for  the Green's function
\begin{equation}
  \label{GN}
  G^{(N)}_{\la }(x,y)=\sum_{\ell\in\bbZ}\big[G_\la\big(x-y+2\ell(2N+1)\big)+G_\la\big(x+y+(2\ell+1)(2N+1)\big)\big],\quad
  x,y\in\bbZ_N,
\end{equation}
with $G_\la$ given by formula  \eqref{012802-23zz}
From \eqref{012802-23} and \eqref{GN} we conclude the following.
\begin{proposition}
  \label{lm021110-24}
  Suppose that there exists $\delta>0$ such that
  \begin{equation}
    \label{031110-24}
    |\la+w|\ge \delta>0,\quad \forall w\in[\om_0^2,\om_0^2+ 4].
  \end{equation}
  Then, there exist constants $C,c>0$, depending only on $\delta$ and
  such that
  \begin{equation}
    \label{041110-24}
    \big|G_{\la}^{(N)}(x,y) \big|\le
    \frac{C}{1+|\la|}\exp\left\{-c|x-y|\right\},\quad x,y\in\bbZ_N.
  \end{equation}
  \end{proposition}

\subsection{Green's function for  $\la+\om^2_0  +i\si(\delta_{-N}
   + \delta_{\ N} )-\Delta$}
\label{secA3}

As before $\Delta$ is the Neumann laplacian on $\bbZ_N$. From \eqref{HN}
we conclude that $H_{\la,\si}^{(N)}(x,y)$ satisfies
\begin{equation} 
\label{011005-21a}
\begin{aligned}
 \big(\la+\om_0^2-\Delta_x ) H_{\la,\si}^{(N)}(x,y) =\delta_{y}(x)-i\si \big(\delta_{-N}(x)+\delta_{N}(x)\big)H_{\la,\si}^{(N)}(x,y)
.
 \end{aligned} \end{equation}
Hence
\begin{equation} 
\label{021205-21b}
 H_{\la,\si}^{(N)}(x,y) =  G_{\la}^{(N)}(x,y)-i\si H_{\la,\si}^{(N)}(-N,y)  G_{\la}^{(N)}(x,-N)
-i\si H_{\la,\si}^{(N)}(N,y) G_{\la}^{(N)}(x,N).
\end{equation}
For $x=-N$ and $x=N$ we get in particular 
 the following
system of equations on  $H_\la^{(N)}(\pm N,y)$
\begin{equation} 
\label{021205-21bb}
\begin{split}
&    H_{\la,\si}^{(N)}(-N,y) \Big(1+i\si    G_{\la}^{(N)}(N,N)\Big)+i\si
H_{\la,\si}^{(N)}(N,y) G_{\la}^{(N)}(-N,N)=  G_{\la}^{(N)}(-N,y)
    \\
    &
  i\si H_{\la,\si}^{(N)}(-N,y)  G_{\la}^{(N)}(N,-N)+
  H_{\la,\si}^{(N)}(N,y)\Big( 1+i\si  G_{\la}^{(N)}(N,N)\Big) =  G_{\la}^{(N)}(N,y).
  \end{split}
\end{equation}

Let
\begin{align*}
 & {\frak Q}_N:=\Big(1+i\si    G_{\la}^{(N)}(N,N)\Big)^2 
    +\si^2\Big( G_{\la}^{(N)}(-N,N)\Big)^2,\\
  &
     {\frak Q}(-N,y):= \Big( 1+i\si  G_{\la}^{(N)}(N,N)\Big)
    G_{\la}^{(N)}(-N,y)  -i\si
    G_{\la}^{(N)}(-N,N) G_{\la}^{(N)}(N,y)  ,\\
  &
     {\frak Q}(N,y):=  \Big( 1+i\si  G_{\la}^{(N)}(N,N)\Big)
    G_{\la}^{(N)}(N,y)  -i\si  G_{\la}^{(N)}(-N,N) G_{\la}^{(N)}(-N,y) ,
\end{align*}
Note that if $\la\in \mathbb R$ and $\si\not=0$ we have
\begin{align}
  \label{030309-24}
{\rm Im}\,{\frak Q}_N=2\si
   G_{\la}^{(N)}(N,N)=2\si\sum_{j=0}^{2N}\frac{\psi_j^2(0)}{\la+\om_j^2}\not=0
   . 
\end{align}
We also have ${\frak Q}_N\not=0$   in the case when $\la\in \mathbb C$,
  and $|{\rm Im}\,\la|$ is sufficiently small.

Then,
\begin{equation} 
  \label{011012-22}
      H_{\la,\si}^{(N)}(-N,y)=\frac{ {\frak Q}(-N,y)}{{\frak Q}_N},\quad
    H_{\la,\si}^{(N)}(N,y)=\frac{ {\frak Q}(N,y)}{{\frak Q}_N}
   \end{equation}

Substituting into \eqref{021205-21b} we get
\begin{align} 
  \label{HNxy}
  H_{\la,\si}^{(N)}(x,y) =  G_{\la}^{(N)}(x,y)- 
i\si G_{\la}^{(N)}(x,-N)\frac{ {\frak Q}(-N,y)}{{\frak Q}_N}-i\si G_{\la}^{(N)}(x,N)\frac{{\frak Q}(N,y)}{{\frak Q}_N} .
  \end{align}

\begin{lemma}
  \label{lm011110-24}
  Suppose that $\la\in\mathbb R$ and
  \begin{equation}
    \label{011110-24}
    |\la+w|\ge \delta>0,\quad w\in[\om_0^2,\om_u^2].
  \end{equation}
  Then, there exist constants $C_*,c_*>0$, depending only on $\delta$ and
  such that
  \begin{equation}
    \label{021110-24}
    \big|H_{\la,\si}^{(N)}(x,y) \big|\le
     \frac{C_*}{1+|\la|}\exp\left\{-c_*|x-y|\right\},\quad x,y\in\bbZ_N.
  \end{equation}
  \end{lemma}

\section{Exponential decay of the periodic solution   in $x$}

\subsection{Some auxiliaries}
  % In the present section we assume that $N$ is fixed and finite.
  Suppose that $\tilde q_x^{(L)}(m;\nu )$, $L=0,{1}, 2,\ldots$ are
   determined by the scheme described in
   \eqref{011005-21-0l}--\eqref{022810-24}, see Section
   \ref{sec3.3.1}. Let $c>0$.
   Define, by recurrence
   $$
e_c(x) :=e_c^{\star,1}(x):=e^{-c|x|}
$$
and
 $$
e_c^{\star,L+1}(x):=\sum_{y\in\bbZ}e^{-c|x-y|}e_c^{\star,L}(y),\quad x\in\bbZ.
$$
We also adopt the convention that
$
e_0(x):=\delta_0(x),\quad x\in\bbZ.
$
Let
 \begin{equation}
    \label{031410-24}
\hat e_c(k):=\sum_{x\in\bbZ} e^{-c|x|}
 \exp\left\{-2\pi i kx\right\}=\frac{1-e^{-2c}}{|1-e^{-c}e^{2\pi i
     k}|^2},\quad k\in\bbT.
\end{equation}
 Then,
  \begin{equation}
    \label{021410-24}
\hat e_c^{\star, L}(k):=\sum_{x\in\bbZ} e_c^{\star,
  L}(x)\exp\left\{2\pi i kx\right\}=\Big(\frac{1-e^{-2c}}{|1-e^{-c}e^{2\pi i k}|^2}\Big)^L
\end{equation}
Since
$$
e_c^{\star, L} (x)=\int_0^1 \Big(\frac{1-e^{-2c}}{|1-e^{-c}e^{2\pi i k}|^2}\Big)^L \exp\left\{2\pi i kx\right\}\dd k,
$$
we can easily conclude that
\begin{equation}
  \label{051410-24}
  0\le   e_c^{\star,L}(x)\le \Big(\frac{1+e^{-c}}{1-e^{-c}}\Big)^L,\quad x\in\bbZ.
  \end{equation}
  For any $q$ such that
  $$
  |q| h(c)<1,\quad\mbox{where}\quad  h(c):=\frac{1+e^{-c}}{1-e^{-c}}
  $$
  we can define therefore
\begin{equation}
  \label{071410-24}
(\delta_0-qe_c)^{-1,\star}(x)=\sum_{L=0}^{+\infty}q^Le_c^{\star,L}(x).
\end{equation}
The series is uniform convergent for all  $x\in\bbZ$.
We also have the following.
\begin{lemma}
  \label{lm011410-24}
  For any $c>0$ and $|q|<1/h(c)$ there exist contstants
  $A(c,q),\rho(c,q)>0$, depending only on the indicated parameters,  such that
  \begin{equation}
    \label{011410-24}
0\le     (\delta_0-q e_c)^{-1,\star}(x)\le
A(c,q)\exp\left\{-\rho(c,q)|x|\right\},\quad x\in\bbZ.
    \end{equation}
  \end{lemma}
  \proof
  From \eqref{071410-24} and \eqref{021410-24} we conclude that
 \begin{equation}
    \label{041410-24}
    \widehat{(\delta_0-qe_c)}^{-1,\star}(k)=\frac{1}{1-q\Big( \frac{1-e^{-2c}}{|1-e^{-c}e^{2\pi i k}|^2}\Big)}.
  \end{equation}
  It is an analytic function in $k\in\bbT$. Therefore, see e.g. \cite[p. 27]{katz}, there exist
  constants $A$ and $\rho>0$ such that
  \begin{equation}
    \label{041410-24a}
   0\le (\delta_0-qe_c)^{-1,\star}(x)\le
   A\exp\left\{-\rho|x|\right\},\quad x\in\bbZ.
  \end{equation}
  \qed

   \begin{lemma}
     \label{lm031110-24}
    Suppose that $c_*>0$ is  as in the statement of Lemma
    \ref{lm011110-24}. Then, there exist $C,c_1>0$ such that
     \begin{equation}
       \label{051110-24a}
       |\tilde q_x^{(L)}(m;\nu )|\le \frac{Cc_1^{L}e_{c_*}^{\star, L}(x)}{1+(m\om)^2},\quad x\in\bbZ_N
     \end{equation}
     for all $N=1,2,\ldots$, $L=0,1,\ldots$ and $m\in\bbZ$.
     \end{lemma}
     \proof
     The estimate for $L=0$ follows from \eqref{012302-24} and
     estimate \eqref{021110-24} with the constant
     $$
     C:=C_*\sup_{m\in\bbZ}|\hat F_m|,
     $$
     where $C_*>0$ is  as in the statement of Lemma
    \ref{lm011110-24} {and $e_{c_*}^{\star, 0}(x)=e^{-c_*|x|}$}.
    Suppose that we have shown that
    \begin{equation}
       \label{051110-24L}
       |\tilde q_x^{(L)}(m;\nu )|\le \frac{Cc^{{L}}_1e_{c_*}^{\star, L}(x)}{1+(m\om)^2},\quad x\in\bbZ_N
     \end{equation}
     for some $L$, with the constant $C_1$ to be determined later on.
     Then,  
    $$ 
q_x^{(L)}(t;\nu )=\sum_{m\in\bbZ}\tilde q_x^{(L)}(m;\nu )e^{im\om t}
$$
we have
 \begin{align}
    \label{071110-24}
&|q_x^{(L)}(t;\nu )|\le \sum_{m\in\bbZ}|\tilde q_x^{(L)}(m;\nu )|\le
                        Cc^{{L}}_1 e_{{c_*}}^{\star,L}(x)\sum_{m\in\bbZ}\frac{1}{1+(m\om)^2}.
\end{align}
Therefore
\begin{align}
  \label{012302-24aa}
  &|\tilde v_{x,L}(m) |\le \frac{1}{|\nu|^{L}\theta}\Bigg\{\int_0^{\theta}\Bigg|V'\big(Q_{x,{\rm
    p}}^{(L)}(t;\nu)\big)-V'\big(Q_{x}
    ^{(L-1)}(t;\nu)\big)\Bigg|
    \dd t\notag\\
  &
    +\int_0^{\theta}\Bigg|\nabla U'\big(Q_{x}^{(L)}(t;\nu)\big)-\nabla U'\big(Q_{x}
    ^{(L-1)}(t;\nu)\big)\Bigg|
  \dd t\Bigg\}\notag 
    \le  \frac{\|V\|_{\rm Lip}+2 \|U\|_{\rm Lip}}{\theta} \int_0^{\theta}|q_x^{(L)}(t;\nu
    )|\dd t\\
  &
    \le  Cc_1^{{L}}(\|V\|_{\rm Lip}+2 \|U\|_{\rm Lip} )e_{{c_*}}^{\star,L}(x)\sum_{m\in\bbZ}\frac{1}{1+(m\om)^2}
\end{align}
and, thanks to \eqref{012302-24a} and \eqref{021110-24},
\begin{align}
    \label{012302-24c}
    &  |\tilde  q_x^{(L+1)}(m ;\nu)|\le
      \sum_{y\in\bbZ_N}|H_{-(m\om)^2, \ga m\om}(x,y)||\tilde
      v_{y,L}(m)|  \notag\\
    &
      \le C{C_*}c_1^{{L}}(\|V\|_{\rm Lip}+2 \|U\|_{\rm Lip} )\Big(\sum_{m'\in\bbZ}\frac{1}{1+(m'\om)^2}\Big)
      \frac{{e_{c_*}^{\star,L+1}(x)}}{1+(m\om)^2}
  \\
    &
      =\frac{C{c_1^{L+1}} e_{{c_*}}^{\star,L+1}(x)}{1+(m\om)^2}, \notag
\end{align}
provided we choose
$$
c_1 := 
{C_*} \sum_{m'\in\bbZ}\frac{\|V\|_{\rm Lip}+2 \|U\|_{\rm Lip}}{1+(m'\om)^2} .
$$
Thus
     \eqref{051110-24a}  follows.
     \qed

     % \begin{theorem}
%        \label{thm011410-24}
%        Suppose that $C,c_1$  are as in the statements of Lemma
%        \ref{lm031110-24}  and 
%        Theorem \ref{main:thm}. Then, assuming that
%        \begin{equation}
%          \label{081410-24}
%        |\nu|<\nu_0\wedge C_1 ,\quad \mbox{where}\quad C_1:=c_1^{-1},
%        \end{equation}
%        there exist $A,\rho>0$ such that
%        \begin{equation}
%          \label{091410-24}
%          \begin{split}
%             &        |q_{x,p}(t;\nu)|\le A\exp\left\{-\rho|x|\right\},\quad
%             t\in[0,\theta],\, t\in[0,\theta] \quad\mbox{and}\\
%             &
%            \int_0^{\theta}p^2_{x,p}(t;\nu)\dd t\le A\exp\left\{-\rho|x|\right\},\quad
%             \,x\in\bbZ_N
% \end{split}
% \end{equation}
% for all $N=1,2,\ldots$.
% \end{theorem}
     \subsection{Proof of Theorem \ref{thm:decay}}
     \label{secC.3}
We have
\begin{align*}
  q_{x,p}(t;\nu)=\sum_{m\in\bbZ} \tilde q_{x,p}(m;\nu)\exp\left\{2\pi
  i m\om t\right\},
\end{align*}
where
$$
\tilde q_{x,p}(m;\nu)=\sum_{L=0}^{+\infty}\tilde q^{(L)}_{x,p}(m;\nu)\nu^L.
$$
Using estimate \eqref{051110-24a} we conclude that
\begin{align*}
&|\tilde q_{x,p}(m;\nu)|\le \frac{C}{C_1(1+(m\om)^2)}\sum_{L=0}^{+\infty} (|\nu|
                 C_1)^{L}e_{c_*}^{\star, L}(x)\\
  &
    =\frac{C}{C_1(1+(m\om)^2)}\Big(\delta_0-(|\nu| C_1)e_{c_*}\Big)^{-1,\star}(x).
\end{align*}
This estimate together with \eqref{011410-24} imply the first bound of
\eqref{exp:decay}.
The second bound follows from the fact that
$$
\frac{1}{\theta}\int_0^{\theta}p^2_{x,p}(t;\nu)\dd t=\sum_{m\in\bbZ} (m\om)^2|\tilde q_{x,p}(m;\nu)|^2.
$$
\qed

\section{The existence of a periodic solution
  for a finite anharmonic
  chain. Proof of Theorem \ref{thm011010-24}}

\label{sec6}

Using the notation of Section \ref{sec4} 
for any $\theta$-periodic $\Omega_N$-valued function  $ {\bf q} (t) $ we can write
\begin{align*}
  &\int_{-\infty}^te^{A(t-s)}   \mathbb V\big( {\bf q} (s )\big)  \dd
    s
    =\sum_{\ell=0}^{+\infty}\int_{0 }^{\theta}e^{A(s+\ell\theta)}
    \mathbb V\big( {\bf q} (t-s-\ell\theta) \big)  \dd s
  \\
  & =\sum_{\ell=0}^{+\infty}\int_{0}^{ \theta}e^{A(s+\ell\theta)}
    \mathbb V\big( {\bf q} (t-s ) \big)  \dd s
    = \int_{0}^{ \theta}(I-e^{A\theta})^{-1}e^{A s }
    \mathbb V\big( {\bf q} (t-s ) \big)  \dd s .
\end{align*}
Suppose that $t\in[0,\theta]$. Then 
  we can  write the utmost right hand side of the above equality as
  being equal to
  \begin{align*}
   \int_{0}^{ \theta}K(t-s)
    \mathbb V( {\bf q} (s))  \dd s,
\end{align*}
where $K$ is the $\theta$-periodic extension of 
\begin{align*}
  K(s)=\left\{
  \begin{array}{ll}
    (I-e^{A\theta})^{-1}e^{A s  },& s\in[0,\theta/2],\\
    &\\
     (I-e^{A\theta})^{-1}e^{A(\theta+ s)  },& s\in[-\theta/2,0].
  \end{array}
  \right.
  \end{align*}
We conclude the following.
\begin{proposition}
  \label{prop011010-24}
  If $ {\bf q}_{\rm p}(t;\nu)$   a $\theta$-periodic solution of
  \eqref{012805-24x}, then it satisfies the equation
  \begin{equation}
    \label{011010-24b}
    \begin{split}
      \left(
  \begin{array}{c}
    {\bf q}_{\rm p}(t;\nu)\\
   {\bf p}_{\rm p}(t;\nu) 
\end{array}
\right)=-\nu\int_{0}^{\theta}K(t-s)  \mathbb V( {\bf q}_{\rm
  p}(s;\nu))  \dd s+\int_{-\infty}^te^{A(t-s)}  \mathbb
F\Big(\frac{s}{\theta}\Big)\dd s.
    \end{split}
  \end{equation}
  Conversely, if $ {\bf q}(t;\nu)$  is a $\theta$-periodic solution of
  \eqref{011010-24b}, then it is a $\theta$-periodic solution of  \eqref{012805-24x}.
\end{proposition}

 \subsubsection*{Proof of Theorem \ref{thm011010-24}}

On the space ${\cal C}_{\rm p}:=C_{\rm p}\Big([0,\theta];\Omega_N\Big)$ of
$\theta$-periodic, $\Omega_N$-valued functions, equipped with the topology of
the uniform convergence on compact intervals,  we introduce the operator
$$
{\cal T}[F](t):=
-\nu\int_{0}^{\theta}K(t-s)  \mathbb V(   \Pi_1F(s))  \dd s+\int_{-\infty}^te^{A(t-s)}  \mathbb
F\Big(\frac{s}{\theta}\Big)\dd s.
$$
Here $\Pi_1 :\bbR^{2(2N+1)}\to \bbR^{2N+1}$ is given by 
$
\Pi_1\left(\begin{array}{c}
   {\bf q}  \\
    {\bf p} 
\end{array}
\right)={\bf q}.
$
We have
${\cal T} :{\cal C}_{\rm p}\to {\cal C}_{\rm p}$. Moreover, ${\cal T}
\big({\cal C}_{\rm p}\big)$ is bounded in ${\cal C}_{\rm p}$ and, since
$t\mapsto K(t)$ is Lipschitz, the set is compact. By the Schauder fixed
point theorem we conclude the existence of a fixed point for ${\cal T}$,
which by virtue of Proposition \ref{prop011010-24} is a periodic
solution to \eqref{012805-24x}.
  \qed
  \medskip

  \section{An application of the implicit function theorem in the
    proof of non-uniqueness of periodic solution}

  \label{appF}

  Recall that $\big(\bar q_x(t;F)\big)_{x\in
  \bbZ_N}$  is a  $\theta=2\pi/\om$-periodic solutions  of
\eqref{012805-24x} corresponding to the single mode forcing \eqref{F-cF}.
 Its time harmonics
$\tilde{\bar q}_x(m;F)$ solve
\begin{equation} 
\label{010907-24z}
\begin{aligned}
   0&=  \big[(\om m)^2-\om_0^2+ \Delta -i\ga  \om m 
    (\delta_{-N}+\delta_{N})\big]\tilde{\bar
  q}_x (m; {\rm  F} ) +{\rm  F}\big(\delta_{1}(m)+\delta_{-1}(m)\big)\delta_{0}(x)\\
  & +\nu\tilde v_{x}(m; {\rm  F})
    , \; \quad (m,x)\in \bbZ\times \bbZ_N,
  \end{aligned} \end{equation}
with 
\begin{equation}
  \label{030810-24z}    
  \begin{split}
&\tilde v_{x}(m; {\rm  F})= \frac{1}{\theta}\int_0^\theta e^{-im\om  t}
v_{x }(t; {\rm  F}  ) \dd t \quad\mbox{and}\\
&
v_{x }(t; {\rm  F} ) :=-V'\big(q_{x,{\rm p}}(t; {\rm  F})\big),
\quad (m,x)\in\bbZ\times\bbZ_N.
\end{split}
\end{equation}
% Let ${\ell}_{2,N}$ be the Hilbert space defined by
% $$
% {\ell}_{2,N}=\big[ \tilde{ f}:=\big(\tilde{
%   f}_x (m ) \big)_{(m,x)\in\bbZ\times\bbZ_N}:\, \tilde{
%   f}_x (-m ) =\tilde{
%   f}^\star_x (m )\quad\mbox{and} \quad\|\tilde{ f}\|^2:=\sum_{m,x}|\tilde{
%   f}_x (m )|^2<+\infty\big].
% $$
We can rewrite the equation in the space ${\ell}_{2,N}$  (see
\eqref{021602-25}) as follows
\begin{equation}
  \label{030810-24y}
  \Phi\Big({\rm F},\tilde{\bar
  q} (\cdot; {\rm  F} ) \Big)=0,
\end{equation}
where $\Phi:\bbR\times \ell_{2,N}\to \ell_{2,N}$ is a mapping given by
   \begin{equation}
    \label{020907-24z}
    \begin{aligned}
& \Phi\big({\rm F},\tilde{
  f} \big)_x(m):=\tilde{
  f}_x (m) -  H^{(N)}_{-(m\om)^2,\ga m\om}(x,0) {\rm F}
\big(\delta_{1}(m)+\delta_{-1}(m)\big)\\
&-  \;     \nu \sum_{y\in\bbZ_N} H^{(N)}_{-(m\om)^2,\ga
  m\om}(x,y)\tilde v(\tilde f)_{y}(m), \quad (m,x)\in\bbZ\times\bbZ_N,\\
&\tilde v(\tilde f)_{x}(m)= \frac{1}{\theta}\int_0^\theta e^{-im\om  t}
V'(f_x(t)) \dd t \quad\mbox{and}\\
&
f_{x }(t) :=\sum_{m\in\bbZ}\tilde f_x(m)e^{i m\om t}
,\quad  t\ge 0.
  \end{aligned}
\end{equation}
For ${\rm F}=0$ equation \eqref{030810-24y} is satisfied by $\tilde{\bar{
    q}}=\big(\tilde{\bar
  q}_x (m; 0 ) \big)$ given by
$$
\tilde{\bar
  q}_x (m; 0 ) =\bar q\delta_{m,0},\quad (m,x)\in\bbZ\times \bbZ_N.
$$
The Fr\'{e}chet derivative  of the right hand side of
\eqref{030810-24y} with respect to the $q$ variable, 
at $(0,\tilde{\bar{
    q}})$  is a linear operator ${\rm L}: {\ell}_{2,N}\to {\ell}_{2,N}$ given by 
\begin{align*}
  & {\rm L}\tilde{ h}_x(m):=\tilde{ h}_x(m) - {\rm K}\tilde{
    h}_x(m),\quad\mbox{where}\\
  &
   {\rm K}\tilde{
    h}_x(m):=     \nu V''(\bar q) \sum_{y\in\bbZ_N}
    H^{(N)}_{-(m\om)^2,\ga m\om}(x,y)\tilde h_y(m)
    , \; \quad (m,x)\in\bbZ\times \bbZ_N.
\end{align*}
It is easy to see that the operator ${\rm K}: {\ell}_{2,N}\to
{\ell}_{2,N}$ is compact. We shall show that its null space is
trivial.
This would imply, see e.g. \cite[Theorem 21.2.6, 238]{lax}, that ${\rm L}$
is onto and its inverse is bounded. Using the implicit function
theorem, see  e.g. \cite[Theorem 4.15.1]{deimling}, we conclude then that
there exists $\bar F_0>0$ such that for any ${\rm F}\in (-\bar F_0, \bar F_0)$ equation \eqref{020907-24z} has a unique solution that gives
rise to a $\theta$ periodic solution  $\bar q_{{\rm p},x}(t;{\rm F})$
of \eqref{012805-24x} that satisfies
$$
\lim_{{\rm F}\to0}\bar q_{{\rm p},x}(t;{\rm F})=\bar q,\quad x\in\bbZ_N.
$$
To see that the kernel of ${\rm L}$ is trivial note that if ${\rm
  L}\tilde{ h}=0$, then
\begin{equation} 
\label{010302-25}
   0=  \big[(\om m)^2-\om_0^2  -\nu V''(\bar q)+ \Delta -i\ga  \om m 
    (\delta_{-N} +\delta_{N})\big]\tilde{h}_x (m ) 
    , \; \quad x\in \bbZ_N.
  \end{equation}
Multiplying both sides pf \eqref{010302-25} by $\tilde{h}_x^\star (m
)$ and summing over $x$ we conclude  that
$$
\sum_{m\in\bbZ}(m\om)^2|\tilde{h}_{\pm N} (m )|^2=0. 
$$
Hence $\tilde{h}_{\pm N} (m )=0$ for $m\not=0$. Arguing as in Section
\ref{sec4.4}, from \eqref{010302-25}
we conclude   that $\tilde{h}_{x} (m )\equiv0$ for all $x\in\bbZ_N$
and $m\not=0$.
For $m=0$ we can multiply both sides of equation \eqref{010302-25} by $\tilde{h}_x^\star (0
)$ and sum over $x$. Since $\om_0^2  +\nu V''(\bar q)>0$ this allows
us to conclude that also in the case $\tilde{h}_{x} (0 )\equiv0$ for
all $x\in\bbZ_N$.

\subsection*{Acknowledgements} We thank David Huse and Abhishek Dhar for
useful discussions. J.L.L. and P.G. wish to express thanks to the
Institute for Advanced Studies for its hospitality where part of this
work was done.

\subsection*{Data availability}

We do not analyse or generate any datasets, because our work proceeds
within a theoretical and mathematical approach.

\subsection*{Conflict of interest statement}

The authors declare no conflicts of interest regarding this manuscript.

\bibliographystyle{amsalpha}

\end{document}